\documentclass[11pt]{revtex4-1}
\usepackage{amsmath, amsthm, amssymb, amsfonts, enumerate}
\usepackage{setspace}
\usepackage{natbib, bbm}
\usepackage{graphicx}

\theoremstyle{plain}

\theoremstyle{definition}

\begin{document}
\title{Quasilinear diffusion coefficients in a finite Larmor radius expansion for ion cyclotron heated plasmas}
\author{Jungpyo Lee}
\affiliation{Massachusetts Institute of Technology, Plasma Science and Fusion Center, Cambridge, MA, USA}
\author{John Wright}
\affiliation{Massachusetts Institute of Technology, Plasma Science and Fusion Center, Cambridge, MA, USA  }
\author{Nicola Bertelli} 
\affiliation{ Princeton Plasma Physics Laboratory, Princeton, NJ, USA}
\author{Erwin F. Jaeger}
\affiliation{XCEL Engineering, Oak Ridge, TN, USA}
\author{Ernest Valeo} 
\affiliation{ Princeton Plasma Physics Laboratory, Princeton, NJ, USA }
\author{Robert Harvey}
\affiliation{CompX, Del Mar, CA, USA}
 \author{Paul Bonoli}
 \affiliation{Massachusetts Institute of Technology, Plasma Science and Fusion Center, Cambridge, MA USA}
\date{\today}

\begin{abstract}
In this paper, a reduced model of quasilinear velocity diffusion by a small Larmor radius approximation is derived to couple the Maxwell's equations and the Fokker planck equation self-consistently for ion cyclotron range of frequency waves in a tokamak. The reduced model ensures the important properties of the full model by Kennel-Engelmann diffusion, such as diffusion directions, wave polarizations, and H-theorem. The kinetic energy change ($\dot{W}$) is used to derive the reduced model diffusion coefficients for the fundamental damping (n=1) and the second harmonic damping (n=2) to the lowest order of the finite Larmor radius expansion. The quasilinear diffusion coefficients are implemented in a coupled code (TORIC-CQL3D) with the equivalent reduced model of dielectric tensor. We also present the simulations of ITER minority heating scenario, in which the reduced model is verified within the allowable errors from the full model results. 
\end{abstract}
\maketitle

\section{Introduction}\label{sec:1}


The ion cyclotron range of frequency (ICRF) waves have been used in a tokamak as a main heating tool \cite{perkins1977heating} or a control tool of MHD phenomena \cite{sauter2002control} (e.g. sawtooth) and turbulent transport \cite{rice2001observations,mantsinen2003application}. In many experiments, the waves are injected to transfer their energy and momentum to plasmas by the fundamental cyclotron resonance with a small population ion species (minority) or by the second harmonic resonance with a major ion species \cite{perkins1977heating}. To estimate the propagation and damping of the waves theoretically, it is necessary to model ion gyro-motion accurately. A reduced model to capture the effect of gyro-motion in the wave-plasma interactions has been developed by assuming a small Larmor radius compared to the wave perpendicular wavelength, which is typically valid in many scenarios of tokamak \cite{brambilla1988local,brambilla1989finite}. The reduced model is used in many numerical codes such as CYRANO \cite{lamalle1994nonlocal}, EVE \cite{dumont2009variational}, PSTELION \cite{vdovin20013d}, and TORIC \cite{brambilla1999numerical}, and it can reduce the computation cost and numerical complexity compared to a full model without the small Larmor radius assumption (e.g. AORSA \cite{Jaeger:POP2001,jaeger2006global}), while including the sufficiently accurate gyro-motion model. In this assumption, the dielectric tensor of plasmas is expanded by a small parameter $k_\perp \rho_i$. Here, $k_\perp$ is the perpendicular wavevector and $\rho_i$ is the ion Larmor radius. The finite Larmor radius (FLR) expansion up to the second order $O((k_\perp \rho_i)^2)$ is sufficient to model the fundamental damping and the second harmonic damping of the fast wave branch \cite{brambilla1988local,brambilla1989finite}. 

A kinetic description of the ICRF wave propagation and damping is important because there is a significant portion of the wave energy is deposited on fast ions. For the kinetic description, the Maxwell's equations are solved for the electric and magnetic fields of the waves and the Fokker-Plank equation is solved for the balance between Coulomb collisions and the particle acceleration due to the wave fields. These two equations should be solved self-consistently, and it is typically obtained by iterating two non-linearly coupled codes. In this iteration process, the quasilinear velocity diffusion coefficients are used to define the acceleration term of the Fokker-Plank equation. The electric and magnetic fields result in velocity space diffusion that can be described by quasilinear theory when the perturbation by the wave is sufficiently small \cite{Cary:PRL1990}. The quasilinear description that represents the average of two linearly perturbed quantities is known to be valid within some acceptable deviation \cite{laval1999controversies, Lee:POP2011}, and the coefficients are proportional to the square of the wave field intensity (or the wave power equivalently). The quasilinear diffusion coefficients were derived in the wavevector $\mathbf{k}$ spectrum space by Kennel and Engelmann (K-E) \cite{Kennel:POF1966}, as summarized in Appendix A. It assumes that the particle trajectory is not perturbed by the wave fields and the magnetic field along the trajectory is homogeneous. As a result, the coefficients do not consider the finite orbit width of particles, which may be important in the low aspect ratio of toroidal geometry \cite{petrov2016fully}. In this paper, for simplicity, we also persist the assumptions of K-E coefficients that are acceptably valid when the inhomogeneity of the magnetic field and the wave power density are not significantly large \cite{laval1999controversies, Lee:POP2011}. 

 The quasilinear diffusion coefficients are evaluated differently in the numerical codes according to their assumptions and formulations. Some advanced model for quasilinear diffusion has been developed by analytically considering the particle trajectory \cite{catto1992quasilinear} and the decorrelation between resonances \cite{lamalle1997radiofrequency} in the toroidal geometry, and numerically evaluating those effects \cite{johnson2006analysis,Harvey2011DC}. The numerical diffusion coefficients are obtained by measuring the diffusion of test particles in a realistic geometry to take account of the finite orbit width and the perturbed orbit. With the assumption of the homogenous magnetic field along the unperturbed trajectory, in the full model without the FLR approximation \cite{jaeger2006global}, the K-E quasilinear diffusion coefficients can be implemented without significant modifications. In the reduced model for the wave solver that does not evaluate $k_\perp$ explicitly, the $k_\perp$ can be approximately estimated by the dispersion relation of a specific branch (e.g. fast wave) to evaluate the K-E coefficients (e.g. TORIC-SSFPQL \cite{brambilla2009advances}). Although the K-E diffusion coefficients in the reduced model code are useful to evaluate the energetic ion tails, they result in an inconsistency between the dielectric tensor in the FLR approximation and the K-E diffusion coefficients without the approximation. 

In this paper, we develop the FLR approximation of the quasilinear diffusion coefficients to be consistent with the dielectric tensor in the reduced model. This approximation of the quasilinear formulation for the FLR expansion is not trivial in terms of several points. Because the numerical formulation to calculate the power absorption is different in each code depending on its assumption and code environment (e.g. geometry, coordinates), the quasilinear diffusion coefficients need to be reformulated to correspond to the power absorption. We formulate the quasilinear coefficients by addressing the following questions: (1) is it possible to use the FLR expansion of the wave power absorption by $\dot{W}$ and $ \mathbf{J}\cdot \mathbf{E}$ for the quasilinear diffusion? (2) is it possible to keep the important characteristics of K-E diffusion (diffusion direction, wave polarization effect, and H-theorem) in the FLR expansion? Here, $\mathbf{J}$ is the plasma current vector, $\mathbf{E}$ is the electric field and $\dot{W}$ is the kinetic energy change which will be defined in Section \ref{sec:2}. We will show that there are some problems with the FLR expansion in terms of the above issues, but the solutions will be suggested in Section \ref{sec:3}.  

 Our quasilinear diffusion coefficients are implemented in a coupled code for the Maxwell equation and the Fokker-Plank equation, TORIC \cite{brambilla1999numerical}-CQL3D \cite{Harvey:1992}, as will be shown in Section \ref{sec:4}. The wave code for the reduced model, TORIC, uses the second order FLR expansion to solve the Swanson-Colestock-Kashuba (SCK) wave equations for the ICRF propagation and damping in a tokamak. It uses a spectral method in both poloidal and toroidal directions, and a cubic finite element method in the radial direction. The second order expansion results in the differential operator of the continuous radial elements, and this numerical formulation results in the distinct characteristics of the reduced model compared to the full model that uses the radial spectral mode for $k_\perp$. In the wave equation in a toroidally symmetric geometry, the toroidal wave spectral modes are decoupled, but the poloidal and radial spectral modes are coupled to each other. For each toroidal spectral mode, the computation time of TORIC solver is about $O(n_r(6n_p)^3)$ while it is about $O((3n_rn_p)^3)$ in the full model code, AORSA. Here $n_p$ is the number of poloidal spectral modes and $n_r$ is the number of radial elements in TORIC or the radial spectral mode in AORSA. In TORIC, the FLR expansion by the SCK approximation reduces the degree of the complexity in the poloidal and radial coupling because only two adjacent radial elements are coupled with the poloidal modes of the element. In spite of the merits of this approach, it is not possible to find the accurate $k_\perp$ due to the missing radial spectral mode, thus TORIC has the inherent limitation to model gyro-motion by the Bessel functions with the argument $k_\perp \rho_i$, as done in the K-E quasilinear diffusion coefficients. In this paper, we suggest the alternative method to model the quasilinear diffusion coefficients for the reduced model without using $k_\perp$. 
 
 The code CQL3D \cite{Harvey:1992,petrov2016fully} solves the bounce-averaged Fokker-Plank equation. It has one radial real space coordinate and two velocity coordinates in the gyro-averaged velocity space. Hence, the quasilinear coefficients in Section \ref{sec:4} are bounce-averaged at each radius. We implement the equivalent FLR expansions in the quasilinear coefficients and the dielectric tensor of the coupled code. Thus, the self-consistent solutions of the wave fields and the distribution functions are obtained both for the minority fundamental damping ($n=1$) and the second harmonic damping ($n=2$) of major ions. Here, $n$ is defined by the range of wave frequency $\omega\simeq n\Omega$, where $\Omega$ is the ion cyclotron frequency.

The rest of the paper is organized as follows. In Section \ref{sec:2}, we explain the kinetic energy change $\dot{W}$ by the ICRF fundamental damping ($n=1$) in the FLR expansion and point out some problems in calculating the quasilinear diffusion coefficients. In Section \ref{sec:3}, some solutions to resolve the problems are suggested and then using the same solution we define the coefficients for the second harmonic damping ($n=2$). In Section \ref{sec:4}, the quasilinear diffusion coefficients and dielectric tensor in the reduced model are implemented in the coupled code, TORIC-CQL3D. Until Section \ref{sec:3}, the formulae are derived non-relativistically for simplicity, but in Section \ref{sec:4} they are expressed relativistically for CQL3D using the relativistic velocity coordinate. Some examples using the codes are shown in Section \ref{sec:5}. Finally, a discussion is given in Section \ref{sec:6}. 

\section{Derivation of $\dot{W}$}\label{sec:2}
In Section \ref{sec:2_1}, we revisit the derivation of the quasilinear diffusion coefficients using $\dot{W}$ in the $k$ spectrum without the FLR approximation. Then, in Section \ref{sec:2_2}, we expand $\dot{W}$ in terms of a small Larmor radius, and explain some problems of calculating the quasilinear coefficients in this approximation.
  
\subsection{$\dot{W}$ without FLR approximation}\label{sec:2_1}

The increase of the kinetic energy density of plasmas due to the energy transfer from RF waves can be described by \cite{brambilla1988local}
\begin{eqnarray}
\int_{-\infty}^{t} \dot{W} dt^\prime&=&q\int d\mathbf{v}\bigg \langle \int_{-\infty}^{t} dt^\prime \{Re[\mathbf{E}(\mathbf{r}^\prime, t^\prime)]\cdot \mathbf{v}^\prime \}\{ Re[\widetilde{f}(\mathbf{v}^\prime,\mathbf{r}^\prime, t^\prime)] \}\bigg\rangle_w,
\end{eqnarray}
where $q$ is the species charge, $\widetilde{f}$ is the perturbed distribution function of the species due to the RF waves and $\left \langle... \right \rangle_w $ indicates the average over a number of wave periods in time and space. Here, $\omega$ is the RF wave frequency, and $\mathbf{v}^\prime$ and $\mathbf{r}^\prime$ denote the velocity and space vector at the past time $t^\prime$, respectively. Using the solution of the linearized Vlasov equation for $\widetilde{f}$ and the time harmonic form for the frequency $\omega$, the energy increase rate (so-called Wdot) is
\begin{eqnarray}
\dot{W} &=&\frac{q^2 \omega}{m}\int d\mathbf{v}  \lim _{\gamma\rightarrow 0} \frac{\gamma}{\omega} Re\bigg [  \int_{-\infty}^{t} dt^\prime e^{-i\omega^*(t-t^\prime)}\left( \mathbf{E}(\mathbf{r}^\prime)^* \cdot {\mathbf{v}^\prime}\right) \nonumber\\
&& \cdot   \int_{-\infty}^{t^\prime} dt^{\prime\prime} e^{i\omega(t-t^{\prime\prime})}\left(\mathbf{E}(\mathbf{r^{\prime\prime}})+\frac{i}{\omega}\mathbf{v}(\mathbf{r^{\prime\prime}}) \times (\nabla \times \mathbf{E}(\mathbf{r^{\prime\prime}}))\right)\cdot \left(\nabla_\mathbf{v^{\prime\prime}} f \right) \bigg ], 
\end{eqnarray}
where $m$ is the mass of the species, $\gamma$ is the damping rate of the wave, and $f$ is the background distribution function. This general expression of the wave damping can be reformulated differently in each numerical code depending on its specific numerical formation and assumption. Using the Fourier spectral representation of the space coordinate, it may be described by
\begin{eqnarray}
\dot{W} &=&\frac{1}{2} Re \bigg [\sum_{\mathbf{k_1}}\sum_{\mathbf{k_2}} e^{i(\mathbf{k_1}-\mathbf{k_2})\cdot \mathbf{r}}(\mathbf{E}_{\mathbf{k_1}} \cdot \mathbf{W_l} \cdot \mathbf{E}_{\mathbf{k_2}}) \bigg],\label{pow1}
\end{eqnarray}
where $ \mathbf{W_l} $ is the resonance kernel and, for example, it is defined in Eq. (11) of \cite{jaeger2006global} for AORSA.

In this paper, we represent the quasilinear diffusion tensor in spherical coordinates, $(v, \vartheta, \phi)$, where $v=\sqrt{v_\perp^2+v_\|^2}$ is the speed, $\vartheta=\arctan(v_\perp/v_\|)$ is the pitchangle, and $\phi$ is the gyroangle. Here, $v_\perp$ and $v_\|$ are the velocity perpendicular and parallel to the static magnetic field, respectively. This coordinate is beneficial to evaluate the quasilinear diffusion tensor because it can reduce the computation time due to the adiabatic invariant, $v$, as will be explained in Section \ref{sec:4}. Then, the quasilinear diffusion coefficients ($\textsf{B}$, $\textsf{C}$, $\textsf{E}$ and $\textsf{F}$) determine the divergence of the flux in velocity space by \cite{kerbel1985kinetic}
 \begin{eqnarray}
 Q(f) =\frac{1}{v^2} \frac{\partial }{\partial v}\left \lbrace {\textsf{B}}\frac{\partial }{\partial v}+\textsf{C} \frac{\partial }{\partial \vartheta} \right \rbrace f  +\frac{1}{v^2 \sin \vartheta} \frac{\partial }{\partial \vartheta}\left \lbrace {\textsf{E}}\frac{\partial }{\partial v}+\textsf{F} \frac{\partial }{\partial \vartheta} \right \rbrace f. 
 \end{eqnarray}
 Because the gyro-averaged quasilinear diffusion coefficients (averaged in $\phi$) are sufficient to model the energy transfer, we define the four coefficients in $Q(f)$ and do not retain the flux in the gyro-phase direction. For the perpendicular momentum transfer, the flux along the gyro-phase direction should be included \cite{Lee:PPCF2012}, but it is not interest of this paper.
 
The energy transfer using the quasilinear diffusion coefficient is
   \begin{eqnarray}
 \dot{W}&=&\int d \mathbf{v} \frac{mv^2}{2} Q(f) = -4\pi\int_0^\infty dv {mv} \int_{0}^{\pi} d\vartheta\sin \vartheta \left(\textsf{B} \frac{\partial f}{\partial v}+\textsf{C}\frac{\partial f}{\partial \vartheta}\right)\label{Wdot1}.
\end{eqnarray}
Using the resonance kernel in Eqs. (\ref{pow1}) and (\ref{Wdot1}), we can evaluate the quasilinear diffusion coefficients for any numerical formulation. Note that the Kennel-Engelmann quasilinear diffusion operator is derived in uniform plasmas with a homogenous static magnetic field, as shown in Appendix A. In a toroidal geometry, we redefine $\dot{W}$ using the local space coordinate $(x,y,z)$ where $x$ and $y$ coordinates are orthogonal in the plane that is perpendicular to the static magnetic field along the $z$ coordinate. The unit vector for $(x,y,z)$ coordinates are $\mathbf{e_x}$, $\mathbf{e_y}$, and $\mathbf{e_\|}$, respectively. For convenience, we define the rotating coordinate $\mathbf{e_+}=(\mathbf{e_x}+i\mathbf{e_y})/\sqrt{2}$ and $\mathbf{e_-}=(\mathbf{e_x}-i\mathbf{e_y})/\sqrt{2}$ and the electric field $E_+=\mathbf{E} \cdot \mathbf{e_+}$ and $E_-=\mathbf{E} \cdot \mathbf{e_-}$ \cite{brambilla1988local}. For example, 
the diagonal components of the quasilinear tensor in the speed direction $v$ is 
\begin{eqnarray}
\textsf{B} &=&\frac{\pi \epsilon \omega_p^2}{ 2 m n_s} \sum_{n} Re \bigg[ { \sum_{\mathbf{k_2}}v_\perp\chi_{\mathbf{k_2} ,n}^*}e^{-i(\mathbf{k_2}\cdot \mathbf{r})}  {\sum_{\mathbf{k_1}}v_\perp\chi_{\mathbf{k_1} ,n}}\delta(\omega-n\Omega-v_\|k_{\| 1})e^{i(\mathbf{k_1}\cdot \mathbf{r})}\bigg] \label{Bfull},
\end{eqnarray}
where $\epsilon$ is the vacuum permittivity, $\Omega$, $\omega_{p}$ and $n_s$ are the gyrofrequency, the plasma frequency and the density of the species, respectively, and $k_{\|}=\mathbf{k}\cdot \mathbf{e_\|}$ is the parallel wavevector. The dirac-delta function is obtained by Plemelj relation \cite{Stix:AIP1992} for $(\omega-n\Omega-v_\|k_{\| })^{-1}$ term in the resonance kernel in Eq. (\ref{pow1}) and the contribution of the wave polarization is determined by the effective potential $\chi_{\mathbf{k} ,n}$ (see Appendix A),
 \begin{eqnarray}
 \chi_{\mathbf{k} ,n}= \frac{1}{\sqrt{2}}E_{\mathbf{k} ,+} J_{n-1} + \frac{1}{\sqrt{2}}E_{\mathbf{k} ,-} J_{n+1}+\frac{v_{\|}}{v_{\perp}}E_{\mathbf{k} ,\|} J_{n},
 \label{chieff}
\end{eqnarray}
where $J_n=J_n(k_\perp\rho_i)$ is the first kind Bessel function for the order n.
The relations between the quasilinear coefficients are determined by the diffusion direction. The diffusion direction is fixed by $G(f)=0$ \cite{Kennel:POF1966,Stix:AIP1992}, where the operator $G$ in Eq. (\ref{KE_diff}) can be also described in spherical coordinates, 
\begin{eqnarray}
G(f)&=&\frac{1}{v}\left( \frac{\partial f}{\partial v}+\left(\frac{v_\|}{v} - \frac{k_{\|} v}{\omega}\right) \frac{1}{v_{\perp}}\frac{\partial f}{\partial \vartheta}\right).
\end{eqnarray}
The operator $G$ results in the Onsager relations of the coefficients \cite{kerbel1985kinetic}:
  \begin{eqnarray}
\textsf{C} &=& \textsf{B}\frac{1 }{v \sin \vartheta} \left[ \cos \vartheta - \frac{k_{\|} v}{\omega}\right], \label{BtoC} \\
\textsf{E} &=& \textsf{B}\frac{1 }{v }\left[ \cos \vartheta - \frac{k_{\|} v}{\omega}\right], \label{BtoE}\\
\textsf{F} &=& \textsf{B}\frac{1 }{v^2 \sin \vartheta}\left[ \cos \vartheta - \frac{k_{\|} v}{\omega}\right]^2 \label{BtoF}.
\end{eqnarray}
These relations are important when we derive the expansion of the quasilinear diffusion coefficients in terms of small Larmor radius in the following sections. The relations should be preserved to fix the diffusion direction in any approximation. Because the relations do not depend on the perpendicular wavevector, $k_\perp$, it can be used in the FLR expansion.

\subsection{$\dot{W}$ in FLR expansion}\label{sec:2_2}
 When the ion Larmor radius is much smaller than the perpendicular wavelength, $\dot{W}$ can be expanded by \cite{brambilla1988local},
 \begin{eqnarray}
\dot{W} &\simeq&\frac{q^2 \omega}{m} \lim _{\gamma\rightarrow 0} \frac{\gamma}{\omega} \int d\mathbf{v}  \left[1+(\mathbf{r}-\mathbf{r_g})\cdot \nabla+\frac{1}{2}(\mathbf{r}-\mathbf{r_g})(\mathbf{r}-\mathbf{r_g}):\nabla\nabla+...\right]\nonumber\\
  &&\times Re\bigg [ 
  \int_{-\infty}^{t} dt^\prime e^{-i\omega^*(t-t^\prime)} \left[1-(\mathbf{r}-\mathbf{r_g})^\prime\cdot \nabla+\frac{1}{2}(\mathbf{r}-\mathbf{r_g})^\prime(\mathbf{r}-\mathbf{r_g})^\prime:\nabla\nabla+...\right] \left( \mathbf{E}(\mathbf{r}^\prime)^* \cdot {\mathbf{v}^\prime}\right) \nonumber\\
&& \cdot   \int_{-\infty}^{t^\prime} dt^{\prime\prime} e^{i\omega(t-t^{\prime\prime})} \left[1-(\mathbf{r}-\mathbf{r_g})^{\prime\prime}\cdot \nabla+\frac{1}{2}(\mathbf{r}-\mathbf{r_g})^{\prime\prime}(\mathbf{r}-\mathbf{r_g})^{\prime\prime}:\nabla\nabla+...\right]\nonumber\\
  &&\times \left(\mathbf{E}(\mathbf{r^{\prime\prime}})+\frac{i}{\omega}\mathbf{v}(\mathbf{r^{\prime\prime}}) \times (\nabla \times \mathbf{E}(\mathbf{r^{\prime\prime}}))\right)\cdot \left(\nabla_\mathbf{v^{\prime\prime}} f \right) \bigg ]\
   \label{WFLR},
\end{eqnarray}
where the SCK approximation is used as in Eq. (23) of \cite{brambilla1988local}. In \cite{brambilla1988local} the electromagnetic contribution is ignored for the Maxwellian distribution function $f=f_M(v)$ (i.e. $(\mathbf{v}\times \mathbf{B}) \cdot \nabla_v f_M=0$), but it is included in Eq. (\ref{WFLR}) for an arbitrary distribution function $f(v, \vartheta)$.
For the small Larmor radius compared to the perpendicular wavelength (i.e. $k_\perp \rho_i \ll1$), we use the Taylor series for the position vector $\mathbf{r}$ from the guiding center position $\mathbf{r_g}$. In this approximation, we expand it by
\begin{eqnarray}
(\mathbf{r}-\mathbf{r_g})^\prime=(\mathbf{r}-\mathbf{r_g})^{(1)\prime}+(\mathbf{r}-\mathbf{r_g})^{(2)\prime}+...,
\end{eqnarray}
where the number in the parenthesis of the superscript denotes the order in the small parameter $k_\perp \rho_i$. The velocity in the lowest order is
  \begin{eqnarray}
  \mathbf{v}^{(0)\prime}=v_\| \mathbf{e_\|}^\prime+\frac{v_\perp}{\sqrt{2}}\left\{e^{i(\phi+\int_t^{t^\prime}\Omega^\prime d\tau)}\mathbf{e_+}^\prime+e^{-i(\phi+\int_t^{t^\prime}\Omega^\prime d\tau)}\mathbf{e_-}^\prime\right\},
    \end{eqnarray}
      where $\phi$ is the gyroangle between $v_\perp$ and $v_x$ in the perpendicular velocity plane at the time $t$. The position vector from the gyro-center in the first order is 
        \begin{eqnarray}
  (\mathbf{r}-\mathbf{r}_g)^{(1)\prime}=-\frac{\mathbf{v}^{(0)\prime} \times \mathbf{e_\|}}{\Omega}=i\frac{v_\perp}{\sqrt{2}\Omega}\left\{e^{i\phi}e^{i\int_t^{t^\prime}\Omega^\prime d\tau}\mathbf{e_+}^\prime-e^{-i\phi}e^{-i\int_t^{t^\prime}\Omega^\prime d\tau}\mathbf{e_-}^\prime\right\}.
    \end{eqnarray}
 
 In Eq. (\ref{WFLR}), the electrostatic contribution is associated with both quasilinear coefficients $\textsf{B}$ and $\textsf{C}$ by
 \begin{eqnarray}
&&\mathbf{E}\cdot \nabla_\mathbf{v} f =E_\perp (\mathbf{e_\perp}\cdot \nabla_\mathbf{v} f)+E_\| (\mathbf{e_\|}\cdot \nabla_\mathbf{v} f) \nonumber\\
&&=\left(E_\perp \frac{v_\perp}{v}+E_\| \frac{v_\|}{v}\right)\frac{\partial f}{\partial v}+\left(E_\perp v_\| -E_\| v_\perp \right)\frac{\partial f}{\partial \vartheta},
\end{eqnarray} 
where $E_\perp(\mathbf{r}^\prime)=\mathbf{E} \cdot (\mathbf{v}^\prime-v_\|^\prime \mathbf{e_\|})/v_\perp^\prime=\{E_+\exp({i(\phi+\int_t^{t^\prime}\Omega^\prime d\tau)})+E_-\exp({-i(\phi+\int_t^{t^\prime}\Omega^\prime d\tau)})\}/\sqrt{2}$. The electromagnetic contribution depends on only the coefficient $\textsf{C}$ that is associated with the pitch-angle direction variation of the distribution function ($\partial f/{\partial \vartheta}$) by
\begin{eqnarray}
&&\left(\frac{i}{\omega}\mathbf{v}(\mathbf{r^\prime}) \times (\nabla \times \mathbf{E}(\mathbf{r^\prime}))\right)\cdot \left(\nabla_\mathbf{v^\prime} f \right)=-\frac{i}{\omega} \left( \nabla \times \mathbf{E}\right)\cdot \left(\mathbf{v} \times \nabla_\mathbf{v^\prime} f \right) \nonumber\\
&&=-\frac{i}{\omega} (\nabla_\| E_\perp-\nabla_\perp E_\|)\frac{\partial f}{\partial \vartheta},
\end{eqnarray}  
where $\nabla_\perp=\{\exp({i(\phi+\int_t^{t^\prime}\Omega^\prime d\tau)})\partial_++\exp({-i(\phi+\int_t^{t^\prime}\Omega^\prime d\tau)})\partial_-\}/{\sqrt{2}}$, $\partial_+=\mathbf{e_+} \cdot \nabla $, $\partial_-=\mathbf{e_-} \cdot \nabla $, and $\nabla_\|=\partial_\|=\mathbf{e_\|} \cdot \nabla $.

For simplicity of the representation and the connection to the coefficients, we can separate the contributions on $\dot{W}$ by each harmonic gyrofrequency ($n$) and by the diffusion directions associated with $\textsf{B}$ and $\textsf{C}$. The $\textsf{B}$ part that is associated with $\partial f/\partial v$ for the ion fundamental damping ($n=1$) is \cite{brambilla1989finite}
      \begin{eqnarray}
\dot{W}^{n=1}_B &\simeq& \frac{q^2 \omega}{m} \lim _{\gamma\rightarrow 0} \frac{\gamma}{\omega} \int d\mathbf{v} v\frac{\partial f}{\partial v} \bigg\{\frac{1}{2}\bigg | \int_0^{\infty} d\tau e^{i\int_0^\tau (\omega-\Omega) d\tau} E_+^\prime \frac{v_\perp^\prime}{v}\bigg|^2 \nonumber \\ 
&-&Re \bigg[ \bigg ( \int_0^{\infty} d\tau^* e^{i\int_0^\tau (\omega^*-\Omega) d\tau} E_+^{\prime*} \frac{v_\perp^\prime}{v}\bigg ) \bigg\{\bigg (\int_0^{\infty} d\tau e^{i\int_0^\tau (\omega-\Omega) d\tau} i \frac{v_\|}{v} \frac{{v^{\prime}_{\perp}}}{\Omega}\partial_+ E_\|^{\prime} \bigg)  \nonumber \\
&&+ \bigg (  \int_0^{\infty} d\tau e^{i\int_0^\tau (\omega-\Omega) d\tau} \frac{{v^{\prime}_{\perp}}^2}{2\Omega^2}((\partial_+\partial_-+\partial_-\partial_+)E_+^{\prime} - \partial_+^2 E_-^{\prime}) \frac{v_\perp^\prime}{v}\bigg)\bigg\} \bigg] \bigg\} \nonumber \\ 
&&- \frac{q^2 \omega}{m} \lim _{\gamma\rightarrow 0} \frac{\gamma}{\omega}\bigg\{ (\partial_+\partial_-+\partial_-\partial_+)\int d\mathbf{v}\frac{{v^{\prime}_{\perp}}^2}{2\Omega^2} v\frac{\partial f}{\partial v}\bigg |\int_0^{\infty} d\tau e^{i\int_0^\tau (\omega-\Omega) d\tau} E_+^\prime\frac{v_\perp^\prime}{v}\bigg|^2 \bigg \} \label{Wdotfun},
 \end{eqnarray}
 where the term in the first line on the right hand side of Eq. (\ref{Wdotfun}) is for the lowest order $O(1)$, the term in the second line is in the first order of $O(k_\perp \rho_i)$, and the remaining terms are in the second order of  $O((k_\perp \rho_i)^2)$. 
 
 The $\textsf{C}$ part that is associated with $\partial f/\partial \vartheta$ for ion fundamental damping ($n=1$) is determined by both electrostatic and electromagnetic parts (i.e. $\dot{W}^{n=1}_{C}=\dot{W}^{n=1}_{C,ES}+\dot{W}^{n=1}_{C,EM}$). The electrostatic part is
       \begin{eqnarray}
\dot{W}^{n=1}_{C,ES} &\simeq& \frac{q^2 \omega}{m}  \lim _{\gamma\rightarrow 0} \frac{\gamma}{\omega}  \int d\mathbf{v} v Re\bigg[ \bigg (  \int_0^{\infty} d\tau e^{i\int_0^\tau (\omega^*-\Omega) d\tau} E_+^{\prime*} \frac{v_\perp^\prime}{v}\bigg )  \bigg\{\bigg ( \int_0^{\infty} d\tau e^{i\int_0^\tau (\omega-\Omega) d\tau}   E_+^{\prime*} {v_\|^\prime}\frac{\partial f}{\partial \vartheta}\bigg) \nonumber \\ 
&&+\bigg ( \int_0^{\infty} d\tau e^{i\int_0^\tau (\omega-\Omega) d\tau} i  \frac{{v^{\prime}_{\perp}}}{\Omega}\partial_+ E_\|^{\prime} v_\perp\frac{\partial f}{\partial \vartheta} \bigg)  \nonumber \\
&&+ \bigg (  \int_0^{\infty} d\tau e^{i\int_0^\tau (\omega-\Omega) d\tau}  \bigg (\frac{{v^{\prime}_{\perp}}^2}{2\Omega^2}((\partial_+\partial_-+\partial_-\partial_+)E_+^{\prime} - \partial_+^2 E_-^{\prime}){v_\|^\prime}\frac{\partial f}{\partial \vartheta}\bigg)\bigg\} \bigg]  \nonumber \\ 
&&- \frac{q^2 \omega}{m}  \lim _{\gamma\rightarrow 0} \frac{\gamma}{\omega}\bigg\{ (\partial_+\partial_-+\partial_-\partial_+)\int d\mathbf{v} v Re\bigg[ \bigg ( \int_0^{\infty} d\tau^* e^{i\int_0^\tau (\omega^*-\Omega) d\tau} E_+^{\prime*} \frac{v_\perp^\prime}{v}\bigg ) \nonumber \\
&&\times \bigg (  \int_0^{\infty} d\tau e^{i\int_0^\tau (\omega-\Omega) d\tau}   E_+^{\prime*} {v_\|^\prime}\frac{\partial f}{\partial \vartheta}\bigg) \bigg] \bigg \} \label{Wces},
 \end{eqnarray}
 and the electromagnetic part is
      \begin{eqnarray}
\dot{W}^{n=1}_{C,EM} &\simeq&\frac{q^2 \omega}{m}  \lim _{\gamma\rightarrow 0} \frac{\gamma}{\omega} \int d\mathbf{v} v Re\bigg[ \bigg ( \int_0^{\infty} d\tau^* e^{i\int_0^\tau (\omega^*-\Omega) d\tau} E_+^{\prime*} \frac{v_\perp^\prime}{v}\bigg )  \bigg\{\bigg ( \int_0^{\infty} d\tau e^{i\int_0^\tau (\omega-\Omega) d\tau}  (\partial_\| E_+^{\prime}- \partial_+ E_\|^{\prime})\frac{\partial f}{\partial \vartheta}\bigg) \nonumber \\
&&+\left(\int_0^{\infty} d\tau e^{i\int_0^\tau (\omega-\Omega) d\tau}  i \frac{{v^{\prime}_{\perp}}}{\Omega} (\partial_\| E_+^{\prime}- \partial_+ E_\|^{\prime})\frac{\partial f}{\partial \vartheta}\right) \nonumber \\
&&+\left(\int_0^{\infty} d\tau e^{i\int_0^\tau (\omega-\Omega) d\tau} \frac{{v^{\prime}_{\perp}}^2}{2\Omega^2}\left((\partial_+\partial_-+\partial_-\partial_+)(\partial_\| E_+^{\prime}- \partial_+ E_\|^{\prime}) - \partial_+^2 (\partial_\| E_-^{\prime}- \partial_- E_\|^{\prime})\right) \frac{\partial f}{\partial \vartheta}\right)\bigg\} \bigg]  \nonumber \\ 
&&-\frac{q^2 \omega}{m}  \lim _{\gamma\rightarrow 0} \frac{\gamma}{\omega}\bigg\{ (\partial_+\partial_-+\partial_-\partial_+)\int d\mathbf{v} v \frac{{v^{\prime}_{\perp}}^2}{2\Omega^2} Re\bigg[ \bigg (\int_0^{\infty} d\tau^* e^{i\int_0^\tau (\omega^*-\Omega) d\tau} E_+^{\prime*} \frac{v_\perp^\prime}{v}\bigg ) \nonumber \\
&&\times\bigg ( \int_0^{\infty} d\tau e^{i\int_0^\tau (\omega-\Omega) d\tau} (\partial_\| E_+^{\prime}- \partial_+ E_\|^{\prime})\frac{\partial f}{\partial \vartheta}\bigg)\bigg ]\bigg\}, \label{Wcem}
 \end{eqnarray}
where the first line is for the lowest order $O(1)$, the second line is for the first order of $O(k_\perp \rho_i)$, and the remaining is for the second order of  $O((k_\perp \rho_i)^2)$ in both Eq. (\ref{Wces}) and (\ref{Wcem}).

Before formulating the quasilinear diffusion coefficients using the $\dot{W}^{1}_B$ and $\dot{W}^{1}_C$, we note three important problems that occur in the expansion of $\dot{W}$: \\
(1) The coefficient $\textsf{B}$ needs to be positive-definite so that the energy is transferred from the waves to plasmas for the Maxwellian equilibrium distribution and the H-theorem by K-E coefficients is guaranteed \cite{Kennel:POF1966}. However, only the lowest order term in $\dot{W}^{1}_B$ guarantees the positive definite property because of the square form. This problem is also explained in \cite{brambilla1989finite}.\\
(2) The zero order in $\dot{W}^{1}_C$ in the first line has the both contribution from $E_+$ and $E_\|$, while it is only determined by $E_+$ in the Kennel-Engelmann form in the lowest order \cite{Kennel:POF1966}. It results in the different contribution from the wave polarization even in the lowest order. In $\chi_{\mathbf{k} ,n=1}$ of Eq. (\ref{chieff}) for the K-E coefficient, only $E_{\mathbf{k} ,+} J_{0}$ results in the zero order contribution.\\
(3) The relation between $\textsf{B}$ and $\textsf{C}$ is different from the relation between K-E form of Eq. (\ref{BtoC}), which results in the different direction of diffusion.

In the next section, we will resolve these three problems to derive the quasilinear diffusion coefficients correctly in the small Larmor radius limit.

\section{quasilinear diffusion}\label{sec:3}
 \subsection{Selection by the lowest order}\label{sec:3_1}
  A solution of the first problem described in the Section \ref{sec:2} is to retain only the lowest order in the quasilinear diffusion. The lowest order for the fundamental damping ($n=1$) is the zero order in the FLR expansion $O(1)$, while the lowest order for the second harmonic damping ($n=2$) is the second order $O((k_\perp \rho_i)^2)$. Keeping the lowest order is sufficient unless the parameter $k_\perp \rho_i$ of the fast ions is too large, as will be shown in Section \ref{sec:5}. The parameter $k_\perp \rho_i$ is determined by the wave power density, the ion density and mass, and the static magnetic field strength.  
  
  Another advantage of selecting only the lowest order term is the compatibility with the dielectric tensor for the current. The current can be derived in FLR expansion \cite{brambilla1989finite} by
  \begin{eqnarray}
  \mathbf{J}(\mathbf{r})&\simeq& \sum_s\frac{q^2 }{m}\int d \mathbf{u} \left[1+(\mathbf{r}-\mathbf{r_g})\cdot \nabla+\frac{1}{2}(\mathbf{r}-\mathbf{r_g})(\mathbf{r}-\mathbf{r_g}):\nabla\nabla\right]\nonumber\\
  &&\times\bigg\{ \mathbf{v} \left[1-(\mathbf{r_g}-\mathbf{r_g}^\prime)_\perp\cdot \nabla_\perp\right]\int_{-\infty}^{t}dt^\prime   e^{(i \omega(t- t^\prime))} \left[1-(\mathbf{r}-\mathbf{r_g})^\prime\cdot \nabla+\frac{1}{2}(\mathbf{r}-\mathbf{r_g})^\prime(\mathbf{r}-\mathbf{r_g})^\prime:\nabla\nabla\right] \nonumber \\
  &&\times \left(\mathbf{E}(\mathbf{r^\prime})+\frac{i}{\omega}\mathbf{v}(\mathbf{r^\prime}) \times (\nabla \times \mathbf{E}(\mathbf{r^\prime}))\right)\cdot \left(\nabla_\mathbf{v^\prime} f \right)\bigg\} \label{J_r2}.
\end{eqnarray}
where $\sum_s$ is the summation over the species.
In the finite Larmor radius approximation, 
\begin{eqnarray}
  \mathbf{J}(\mathbf{r})=\left[\mathbf{J}^{(0)}+\mathbf{J}^{(1)}+\mathbf{J}^{(2)}+...\right]e^{(-i \omega t)},
  \end{eqnarray}
the zero Larmor radius current is
\begin{eqnarray}
\mathbf{J}^{(0)}(\mathbf{r})=  \sum_s\frac{q^2 }{m} \int d \mathbf{v}  \mathbf{v}^{(0)\prime}\int_{-\infty}^{t}dt^\prime   e^{(i \omega(t- t^\prime))}\left(\mathbf{E}(\mathbf{r^\prime})+\frac{i}{\omega}\mathbf{v}^{(0)\prime} \times (\nabla \times \mathbf{E}(\mathbf{r^{\prime}}))\right)  \cdot \nabla_\mathbf{v^\prime} f (\mathbf{v}^{(0)\prime}).
  \end{eqnarray}
  The $n=1$ fundamental damping is determined by the $\mathbf{e_+}$ in the lowest order. After gyro-average in $\phi$, it can be described by the non-local operators $\hat{L}$, $\Delta\hat{L}_1$ and $\Delta\hat{L}_2$,
\begin{eqnarray}
 \mathbf{e_+} \cdot \left(  \mathbf{E}+\frac{\mu_0 i}{\omega}\mathbf{J}^{(0)}\right)=(\hat{L}+\Delta\hat{L}_1)E_+ +\Delta\hat{L}_2E_\|, \label{Ltot}
  \end{eqnarray}
  where $\mu_0$ is the vacuum permeability and the integral operator $\hat{L}$ \cite{brambilla1999numerical} is
  \begin{eqnarray}
  \hat{L}E_+&=&E_+(\mathbf{r})- \sum_s\frac{\omega_{p}^2}{n_s\omega^2}\int d^3v ({\mathbf{v}^{(0)}\cdot\mathbf{e_+}^*}) ( \nabla_\mathbf{v} f \cdot \mathbf{e_+})\left(-i\omega \int_{-\infty}^t dt^\prime  e^{i\int^t_{t^\prime}(\omega-\Omega^\prime) d\tau}{E_+}^\prime\right) \nonumber\\
  &=&E_+(\mathbf{r})- \sum_s\frac{\omega_{p}^2}{n_s\omega^2}\int d^3v \frac{v_\perp}{2}   \frac{\partial f}{\partial v_\perp} \left(-i\omega \int_{-\infty}^t dt^\prime  e^{i\int^t_{t^\prime}(\omega-\Omega^\prime) d\tau}{E_+}^\prime\right) \label{Lop}.
    \end{eqnarray}
        
The electromagnetic contribution to the integral operator are $\Delta\hat{L}_1$ and $\Delta\hat{L}_2$,
  \begin{eqnarray}
  \Delta\hat{L}_1E_+&=&\sum_s \frac{\omega_{p}^2}{n_s\omega^2}\int d^3u \frac{ v_\perp }{2}  \frac{i}{\omega}  \frac{\partial f}{\partial \vartheta } \left(-i\omega \int_{-\infty}^t dt^\prime  e^{i\int^t_{t^\prime}(\omega-\Omega^\prime) d\tau}\partial_\|{E_+}^\prime\right), \label{L1plus} \\
    \Delta\hat{L}_2E_\|&=&\sum_s \frac{\omega_{p}^2}{n_s\omega^2}\int d^3u \frac{v_\perp }{2}  \frac{i}{\omega}  \frac{\partial f}{\partial \vartheta } \left(i\omega \int_{-\infty}^t dt^\prime  e^{i\int^t_{t^\prime}(\omega-\Omega^\prime) d\tau}\partial_+{E_\|}^\prime\right). 
    \end{eqnarray}
Because of the cancelation by full FLR terms as will be shown in the next subsection, the lowest order of $\dot{W}^{n=1}_{C,EM}$ has only the term with $E_+$ in the first line of Eq. (\ref{Wcem}), and the $ \Delta\hat{L}_2E_\|$ only contributes to the higher order $O(k_\perp \rho_i)$. Then, to the lowest order, one can prove that  

 \begin{eqnarray}
\dot{W}^{n=1(0)} &=&\frac{1}{2}Re({\mathbf{E}^*\cdot \mathbf{J}}^{n=1(0)}) \nonumber \\
&=&\frac{\omega}{2\mu_0} Re\bigg[\sum_{\mathbf{k}_1} \sum_{\mathbf{k}_2} E_{+}^*(\mathbf{k}_2)e^{-i\mathbf{k}_2 \cdot \mathbf{r}} \ Im (\hat{L}+ \Delta\hat{L}_1) E_{+}(\mathbf{k}_1)e^{i\mathbf{k}_1 \cdot \mathbf{r}} \bigg]\label{WFLR0_n1},
\end{eqnarray}
  where we used the SCK approximation, giving the electric field for the trajectory along the static magnetic field by $\mathbf{E}^\prime=\mathbf{E}\left(\mathbf{r}-\int^t_{t^\prime} v_\| \mathbf{e_\|}\prime d\tau\right)$ and
       \begin{eqnarray}
      \nabla_{\mathbf{v}^\prime} f \cdot \mathbf{e_+}^\prime\simeq (\nabla_\mathbf{v} f \cdot \mathbf{e_+}) e^{i\int_t^{t^\prime}\Omega^\prime d\tau},\nonumber\\
      \nabla_{\mathbf{v}^\prime} f \cdot \mathbf{e_-}^\prime\simeq (\nabla_\mathbf{v} f \cdot \mathbf{e_-}) e^{-i\int_t^{t^\prime}\Omega^\prime d\tau}.
          \end{eqnarray}
          
 In the general relation, there exists the contribution of kinetic flux making the difference between $\dot{W}$ and $ \langle\mathbf{E}\cdot \mathbf{J}  \rangle_w $,
\begin{eqnarray}
 \langle \mathbf{E}\cdot \mathbf{J} \rangle_w =\dot{W} + \nabla \cdot \mathbf{T},
\end{eqnarray}
where the kinetic flux $\mathbf{T}$ has been derived in many studies  \cite{brambilla1988local,smithe1989local}. However, to the lowest order, this kinetic flux contribution vanishes in the $n=1$ damping. Also, the $\dot{W}$ and $\langle\mathbf{E}\cdot \mathbf{J}  \rangle_w $ in Eq. (\ref{WFLR0_n1}) are equivalent to the $\dot{W}$ in Eq. (14) of \cite{jaeger2006global} by making $J_0(k_\perp \rho_i)=1$ and $J_{n>0}(k_\perp \rho_i)=0$ in the FLR approximation. For the iteration between TORIC-CQL3D, this equivalence between $\dot{W}$ and $\langle\mathbf{E}\cdot \mathbf{J}  \rangle_w $ in the lowest order makes the numerical implementation much simpler, and we can use the existing implementation for non-Maxwellian dielectric tensor in the lowest order \cite{Phillips2005effects, Bertelli2017full}.
    
\subsection{Cancelation by full FLR expansion}\label{sec:3_2}
The second and third problems described in the Section \ref{sec:2} can be resolved by considering the cancelations of the terms using the summation of all FLR expansions, as described in Chapters 10 and 17 of \cite{Stix:AIP1992}. In this subsection, we explain the cancellation, which can be applicable for both forms in $\dot{W}^{n=1(0)}_B+\dot{W}^{n=1(0)}_C$ and $(\hat{L}+\Delta\hat{L_1})E_++ \Delta\hat{L}_2E_\|$. The methods of cancellation in both forms are the same, so we only derive it for the latter form.

We note two facts in the form of $(\hat{L}+\Delta\hat{L_1})E_++ \Delta\hat{L}_2E_\|$. First, the operator $(\hat{L}+\Delta\hat{L_1})$ has the operator $U$,
\begin{eqnarray}
U &\equiv&   \left( \frac{\partial f}{\partial v_\perp}+\frac{k_\|}{\omega} \frac{\partial f}{\partial \vartheta}\right)=v_\perp G, \label{U_dql}
\end{eqnarray}
where $k_\|$ comes from $\partial_\|$ in Eq. (\ref{L1plus}). The operator $U$ guarantees the diffusion direction of K-E coefficients towards $G=0$. Secondly, the $\Delta\hat{L}_2E_\|$ term cancels with other contributions of $E_\|$ (on $P$ operator in Stix notation \cite{Stix:AIP1992,brambilla1999numerical}) by
\begin{eqnarray}
&&e^{i\lambda \sin {\phi}}  E_\| \frac{\partial f}{\partial v_\|}+\cos{\phi} e^{i\lambda \sin {\phi}}\frac{i}{\omega } \nabla_\perp E_\| \frac{\partial f}{\partial \vartheta}
=E_\| \sum_n e^{in\phi} J_n(\lambda)\left[\frac{\partial f}{\partial v_\|}- \frac{n}{\lambda}\frac{k_\perp }{\omega}\frac{\partial f}{\partial \vartheta}\right]  \nonumber \\
&=&E_\| \sum_n e^{in\phi} J_n(\lambda) \left[\left(1- \frac{n\Omega}{\omega}\right) \frac{1}{v_\perp}\frac{\partial f}{\partial \vartheta}+ \frac{v_\|}{v_\perp}\frac{\partial f}{\partial v_\perp}\right]\nonumber\\
&=&E_\| \sum_n e^{in\phi} J_n(\lambda)\left[\frac{v_\|}{v_\perp} U+\frac{\omega-k_\|v_\|-n\Omega}{\omega}\left(\frac{\partial f}{\partial v_\|}-\frac{v_\|}{v_\perp} \frac{\partial f}{\partial v_\perp}\right)\right],\label{cancel1}
\end{eqnarray}
where $\lambda=k_\perp \rho_i$.
 Because the term for the resonance condition $\omega-k_\|v_\|-n\Omega$ also exists in the denominator of the operators $\Delta\hat{L}_2$ and $P$, the second term in the last line of Eq. (\ref{cancel1}) vanishes for any distribution function $f$. The remaining term depends on the $U$ operator and it is in the higher order of $O(k_\perp \rho_i)$ for n=1 because of $J_1(\lambda)\sim k_\perp \rho_i/2$. Here, when we derive Eq. (\ref{cancel1}), we utilize the Bessel function expansion for the sinusoid phase in Eq. (\ref{Bess00}) and the following Bessel function identities are used for the cancellation
 \begin{eqnarray}
\sum_{n=-\infty}^{\infty} nJ_n^2(\lambda)=0, \;\;\;\sum_{n=-\infty}^{\infty} J_n\frac{dJ_n(\lambda)}{d\lambda}=0,\;\;\;\sum_{n=-\infty}^{\infty} J_n^2(\lambda)=1.
\end{eqnarray} 
Here, note that the full summation over $n$ number is required in the cancellation. The small contributions from the higher orders $n>1$ accumulate for the cancelation in the lower order.

Because the operator $U$ only remains for both $(\hat{L}+\Delta\hat{L_1})$ and $\Delta\hat{L}_2E_\|$, the diffusion direction in the relations of Eq. (\ref{BtoC}-\ref{BtoF}) still holds. Also, to the lowest order, the coefficient $\textsf{B}$ for n=1 is only determined by $\hat{L}$, giving
\begin{eqnarray}
\textsf{B} &=& \frac{\pi \epsilon \omega_p^2}{ 4m n_s} Re \bigg[ { \sum_{\mathbf{k_2}} {v_\perp E_{+,\mathbf{k_2}}^*}e^{-i(\mathbf{k_2}\cdot \mathbf{r})}}  {\sum_{\mathbf{k_1}} v_\perp E_{+,\mathbf{k_1}}}\delta(\omega-\Omega-v_\|k_{\| 1})e^{i(\mathbf{k_1}\cdot \mathbf{r})}\bigg], \label{Bn_1}
\end{eqnarray}
which is the FLR approximation of Eq. (\ref{Bfull}).
   
    \subsection{Second harmonic damping when $\omega\sim 2\Omega$}\label{sec:3_3}
As was done for n=1, we can select only the lowest order FLR contributions of $\dot{W}$ to derive the quasilinear diffusion coefficient for $n=2$. The B part of the $\dot{W}$ for $n=2$ in the lowest order that is associated with ${\partial f}/{\partial v}$ is
            \begin{eqnarray}
            \dot{W}^{n=2}_B &\simeq& \frac{q^2 \omega}{m} \lim _{\gamma\rightarrow 0} \frac{\gamma}{\omega}   \int d\mathbf{v} v\frac{\partial f}{\partial v} \frac{v_{\perp}^2}{\Omega^2}\bigg |\int_0^{\infty} d\tau e^{i\int_0^\tau (\omega-2\Omega) d\tau} \partial_+ E_+^\prime \frac{v_\perp^\prime}{v} \bigg|^2. \label{Wn2}
   \end{eqnarray}
  As shown in Eq. (\ref{Wn2}), the dominant term for $n=2$ is in $O((k_\perp \rho_i)^2)$ \cite{brambilla1988local}. The $C$ part that is associated with $\partial f/\partial \vartheta$ for $n=2$ is determined by both electrostatic and electromagnetic parts (i.e. $\dot{W}^{n=2}_{C}=\dot{W}^{n=2}_{C,ES}+\dot{W}^{n=2}_{C,EM}$). To the lowest order, the electrostatic part is
       \begin{eqnarray}
\dot{W}^{n=2}_{C,ES} &\simeq& \frac{q^2 \omega}{m}\lim _{\gamma\rightarrow 0} \frac{\gamma}{\omega}  \int d\mathbf{v} v  \frac{v_{\perp}^2}{\Omega^2} Re\bigg[ \bigg (  \int_0^{\infty} d\tau e^{i\int_0^\tau (\omega^*-2\Omega) d\tau} \partial_- E_+^{\prime*} \frac{v_\perp^\prime}{v} \bigg ) \nonumber \\
&&\times  \bigg (  \int_0^{\infty} d\tau e^{i\int_0^\tau (\omega-2\Omega) d\tau} \partial_+ E_+^\prime {v_\|^\prime}\frac{\partial f}{\partial \vartheta}\bigg) \bigg],
   \end{eqnarray}
where $ \partial_+^*= \partial_-$ is used. The electromagnetic part is
       \begin{eqnarray}
\dot{W}^{n=2}_{C,EM} &\simeq& \frac{q^2 \omega}{m} \lim _{\gamma\rightarrow 0} \frac{\gamma}{\omega}  \int d\mathbf{v} v  \frac{v_{\perp}^2}{\Omega^2} Re\bigg[ \bigg (\int_0^{\infty} d\tau e^{i\int_0^\tau (\omega^*-2\Omega) d\tau} \partial_- E_+^{\prime*} \frac{v_\perp^\prime}{v} \bigg ) \nonumber \\
&&\times  \bigg (  \int_0^{\infty} d\tau e^{i\int_0^\tau (\omega-2\Omega) d\tau} \partial_+ (\partial_\| E_+^{\prime}- \partial_+ E_\|^{\prime})\frac{\partial f}{\partial \vartheta}\bigg) \bigg].
   \end{eqnarray}

In the SCK approximation, the ion current $\mathbf{J}^{(2,2)}$ retains the term resonant at $\omega=2\Omega$ \cite{brambilla1999numerical}, and it is obtained by the operators, $\hat{\lambda}^{(2)}$, $\Delta \hat{\lambda}_1^{(2)}$, and $\Delta \hat{\lambda}_2^{(2)}$,
      \begin{eqnarray} \mathbf{e_+} \cdot \left( 
\frac{\mu_0 i}{\omega}{J}_{+,B}^{n=2, (2)}\right)&=& \frac{c^2}{\omega^2} \bigg\{2 \partial_-(\hat{\lambda}^{(2)}+\Delta \hat{\lambda}_1^{(2)})\partial_+E_+^{\prime}+2 \partial_-\Delta \hat{\lambda}_2^{(2)}\partial_+E_\|^{\prime}    \bigg\}  \label{J_2+}
 \end{eqnarray}
 where the operator $\hat{\lambda}^{(2)}$ is \cite{brambilla1999numerical}
   \begin{eqnarray}
\hat{\lambda}^{(2)} {E_+}^\prime &=&\sum_s \frac{\omega_{p}^2}{n_sc^2}\int d^3v \frac{v_\perp^3 }{4\Omega^2}   \frac{\partial f}{\partial v_\perp} \left(-i\omega \int_{-\infty}^t dt^\prime  e^{i\int^t_{t^\prime}(\omega-2\Omega^\prime) d\tau}  {E_+}^\prime\right). 
 \end{eqnarray}
The electromagnetic contribution to the integral operator are $\Delta \hat{\lambda}_1^{(2)}$, and $\Delta \hat{\lambda}_2^{(2)}$,
   \begin{eqnarray}
\Delta \hat{\lambda}^{(2)}_1 {E_+}^\prime &=&\sum_s \frac{\omega_{p}^2}{n_sc^2}\int d^3v \frac{v_\perp^3 }{4\Omega^2}    \frac{i}{\omega}  \frac{\partial f}{\partial \vartheta } \left(-i\omega \int_{-\infty}^t dt^\prime  e^{i\int^t_{t^\prime}(\omega-2\Omega^\prime) d\tau}  {\partial_\| E_+}^\prime\right), \\
\Delta \hat{\lambda}^{(2)}_2 {E_\|}^\prime &=&\sum_s \frac{\omega_{p}^2}{n_sc^2}\int d^3v \frac{v_\perp^3 }{4\Omega^2}   \frac{i}{\omega}  \frac{\partial f}{\partial \vartheta }  \left(i\omega \int_{-\infty}^t dt^\prime  e^{i\int^t_{t^\prime}(\omega-2\Omega^\prime) d\tau}  {\partial_+ E_\|}^\prime\right). 
 \end{eqnarray}
 Because of the cancelation by the full FLR expansions as shown in the previous section, the lowest order contributions for $n=2$ in $O((k_\perp \rho_i)^2)$ result in
   \begin{eqnarray}
\dot{W}^{n=2(2)} &=&\frac{1}{2}Re({\mathbf{E}^*\cdot \mathbf{J}}^{n=2(2)}) \nonumber \\
&=&\frac{\omega}{2\mu_0}Re \bigg[\sum_{\mathbf{k}_1} \sum_{\mathbf{k}_2} E_{+}^*(\mathbf{k}_2)e^{-i\mathbf{k}_2 \cdot \mathbf{r}} \ Im (\hat{\lambda}^{(2)} + \Delta \hat{\lambda}^{(2)}_1) E_{+}(\mathbf{k}_1)e^{i\mathbf{k}_1 \cdot \mathbf{r}} \bigg]\label{WFLR0_2},
\end{eqnarray}
where the integration by parts is used. 
Then, to the lowest order for $n=2$, the diffusion direction in the relations of Eq. (\ref{BtoC}-\ref{BtoF}) still holds and the coefficient $\textsf{B}$  is only determined by $\hat{\lambda}^{(2)}$, giving
\begin{eqnarray}
\textsf{B} &=&\frac{\pi \epsilon \omega_p^2}{ 8 m n_s} Re \bigg[  \frac{v_\perp^2 }{\Omega^2} { \sum_{\mathbf{k_2}} {v_\perp E_{+,\mathbf{k_2}}^*} e^{-i(\mathbf{k_2}\cdot \mathbf{r})}} \partial_-\bigg( {\sum_{\mathbf{k_1}} v_\perp \partial_+ E_{+,\mathbf{k_1}}}e^{i(\mathbf{k_1}\cdot \mathbf{r})}\delta(\omega-2\Omega-v_\|k_{\| 1}) \bigg) \bigg]\label{Bn_2},
\end{eqnarray}
which is equivalent to the $J_1(k_\perp \rho_i)=k_\perp \rho_i/2$ for the small Larmor radius approximation of Eq. (\ref{Bfull}), and the $k_\perp$ is replaced by the operators $ \partial_-$ and $ \partial_+$.

\section{Implementation in TORIC-CQL3D}\label{sec:4}
In this section, we explain a specific numerical code (TORIC-CQL3D) to solve the Maxwell equation and Fokker-Plank equation self-consistently using the quasilinear diffusion in Section \ref{sec:3}. The expansion of the quasilinear diffusion coefficients is consistent with the FLR approximation of the wave equations in TORIC. The quasilinear diffusion coefficients are used as the input data of the Fokker-Plank solver, CQL3D, which uses 1-D radial coordinate and 2-D velocity space ($v$ and $\vartheta$ coordinate). The toroidal coordinate is neglected because of toroidal symmetry and the velocity coordinate in the gyro-angle direction is eliminated by the average over the fast gyro-motion. The poloidal dependency is also eliminated by using the bounce-averaged Fokker-Plank equation, in which the parallel streaming term is eliminated by the average \cite{Harvey:1992}. While $v$ is invariant over the poloidal angle $\theta$ of a flux surface, the velocity pitch angle $\vartheta$ is not. Accordingly, CQL3D uses a distribution function that is defined at a poloidal location (outer-midplane) of each flux surface. The effects of the poloidal finite orbit width is included in the modified version of CQL3D \cite{petrov2016fully}, but it is not considered in this paper. To transfer the quasilinear diffusion coefficients to CQL3D, we need to evaluate the bounce-average of the quasilinear coefficients, which are explained in Section \ref{sec:4_1}. Additionally, the coefficients are also described relativistically, because the CQL3D uses the normalized relativistic velocity coordinate at the outer-midplane, $\mathbf{u}_0=\gamma_r \mathbf{v}_0/c$, where $\gamma_r$ is the relativistic factor, $c$ is the speed of light and the subscript $0$ denotes the value at the outer-midplane. Once we find the solution of distribution function in CQL3D, we need to reevaluate the wave fields in TORIC corresponding to the new distribution function. In this case, we need to use the dielectric tensor for the non-Maxwellian plasmas in TORIC, as implemented in \cite{Phillips2005effects, Bertelli2017full}. In Section \ref{sec:4_2}, we briefly mention the equivalence between the quasilinear diffusion coefficients and the dielectric tensor. 
     
\subsection{Bounce-averaging}\label{sec:4_1}
The bounce average  is defined as $
\langle X \rangle_{b}  = ({1}/{\tau_b}) \oint {d\ell X}/|{v_{\|}}|=({1}/{\tau_b}) \oint {d\theta X}/(|v_{\|}| \mathbf{e_\|}  \cdot \nabla\theta)$ along the particle trajectory, where $\ell$ is the trajectory distance and $\tau_b=d\ell/|{v_{\|}}|$ is the bounce time. Using this definition, the $\textsf{B}$ component of bounce-averaged quasilinear diffusion coefficient for n=1 is 
\begin{eqnarray}
 \langle \textsf{B}\rangle_b &=& \frac{\pi \epsilon \omega_p^2}{ 4 m n_s}\frac{1}{\tau_b} \int_0^{2\pi}\frac{d\theta}{|v_\||\mathbf{e_\|} \cdot \nabla \theta} \sum_{m_1} \sum_{m_2} Re \bigg[e^{i(m_2-m_1)\theta} {E_+}(m_1) \nonumber\\
&&\times \left[\frac{u_\perp^2 c^2}{\gamma_r |k_\||}\delta\left(v_\|-\frac{\omega-\Omega}{k_\|} \right) \right]{E_+}(m_2)\bigg], \label{lambdaB1}
\end{eqnarray}
    where the electric field is decomposed into poloidal spectral modes $\sum_{m} \mathbf{E}(m)\exp({im\theta})$ for a fixed toroidal spectral mode at each radial element. The resonance condition $\omega-\Omega-v_\|{k_\|}=0$ is relativistic using $\gamma_r=(1+u^2)^{1/2}$ and $\Omega=\Omega_s/\gamma_r$ where $\Omega_s$ is the gyrofrequency with the rest mass, giving the elliptic equation \cite{Harvey:1992}
     \begin{eqnarray}
     (u_{\|,res}-u_{\|,t})^2+\frac{u_{\perp,res}^2}{1-n_\|^2}=u_t^2,
      \end{eqnarray}
      where $n_\|=k_\|c/\omega$ is the parallel refractive index and
           \begin{eqnarray}
        \frac{u_{\|,t}}{c}&=&\frac{\Omega_s}{\omega}\frac{n_\|}{1-n_\|^2},\nonumber \\
          \left(\frac{u_{t}}{c}\right)^2&=&\left(\left(\frac{\Omega_s}{\omega}\right)^2-(1-n_\|^2)\right)\frac{1}{(1-n_\|^2)^2}.
          \end{eqnarray}
The numerical representation of Dirac delta function in Eq. (\ref{lambdaB1}) has two options. One is to model the delta function as a kernel in parallel velocity coordinate having the delta function properties (e.g.  rectangular, triangular, or gaussian shape) \cite{jaeger2006global}. For simplicity, the rectangular delta function in $u_{\|0}$ coordinate with a small width $\Delta u_{\|0}$ and a large height $1/\Delta u_{\|0}$ is used in the code. The other option is to pre-evaluate the delta function in the integral in terms of poloidal angle, because the argument of delta function varies along the poloidal angle, and giving the local resonance at a poloidal angle $\theta_{res}$ (i.e. $\delta(v_\|-({\omega-\Omega})/{k_\|})=\delta (\theta-\theta_{res})/(\partial (v_\|-({\omega-\Omega})/{k_\|})/\partial \theta )$) \cite{Wright:IEEE2010}. The latter does not use the model of the delta function kernel, so it can be more accurate theoretically. However due to the numerical error by the negative value of bounce-average, the first option is likely better to produce the accurate value of $\langle \textsf{B}\rangle_b$. 

The quasilinear diffusion coefficients for the second harmonic damping in TORIC use the same differential operator as the plasma current for $n=2$. The ion current $\mathbf{J}^{n=2,(2)}$ retaining the term resonant at $\omega=2\Omega$ is
   \begin{eqnarray}
\frac{\mu_0 i}{\omega}\mathbf{J}^{n=2,(2)}(\mathbf{r})&=& \frac{c^2}{\omega^2} \mathbf{R} \cdot \bigg\{\nabla_\perp((\hat{\lambda}^{(2)} + \Delta \hat{\lambda}^{(2)}_1) \nabla_\perp \cdot (\mathbf{R} \cdot \mathbf{E}_\perp)+i(\hat{\lambda}^{(2)} + \Delta \hat{\lambda}^{(2)}_1) \nabla_\perp \cdot (\mathbf{e}_\| \times \mathbf{R} \cdot \mathbf{E}_\perp))\nonumber \\
&&-(\mathbf{e}_\| \times \nabla_\perp)((\hat{\lambda}^{(2)} + \Delta \hat{\lambda}^{(2)}_1) \nabla_\perp \cdot (\mathbf{e}_\| \times \mathbf{R} \cdot \mathbf{E}_\perp))\nonumber \\
&&-i(\hat{\lambda}^{(2)} + \Delta \hat{\lambda}^{(2)}_1) \nabla_\perp \cdot (\mathbf{R} \cdot \mathbf{E}_\perp))\bigg\}, \label{J2_2_toric}
 \end{eqnarray}
 where the matrix $ \mathbf{R} $ is the reflection matrix with respect to the plane containing the static magnetic field $\mathbf{B}_0$, giving $\mathbf{R}  \cdot \mathbf{E}\mp i \mathbf{e}_\| \times (\mathbf{R}  \cdot \mathbf{E})=2E_\pm\mathbf{e}_\pm$  \cite{brambilla1989finite}.
Then, its corresponding bounce-averaged quasilinear coefficient for $\textsf{B}$ is 
\begin{eqnarray}
 \langle \textsf{B}\rangle_b  &=&\frac{1}{\tau_b} \int_0^{2\pi}\frac{d\theta}{|v_\||\mathbf{e_\|}\cdot \nabla \theta} Re \sum_{m_1} \sum_{m_2}e^{i(m_2-m_1)\theta} \nonumber\\&&\times \mathbf{E}_\perp(m_1)\cdot\mathbf{R} \cdot \bigg\{\nabla_\perp( \bar{ \lambda}^{(2)}\nabla_\perp \cdot (\mathbf{R} \cdot \mathbf{E}_\perp(m_2)) + i  \bar{\lambda}^{(2)}\nabla_\perp \cdot (\mathbf{e}_\| \times \mathbf{R} \cdot \mathbf{E}_\perp(m_2))) \nonumber \\
&&-(\mathbf{e}_\| \times \nabla_\perp)( \bar{ \lambda}^{(2)}\nabla_\perp \cdot (\mathbf{e}_\| \times \mathbf{R} \cdot \mathbf{E}_\perp(m_2))-i   \bar{\lambda}^{(2)}\nabla_\perp \cdot (\mathbf{R} \cdot \mathbf{E}_\perp(m_2)))\bigg\},
\end{eqnarray}
where the redefined operator $\bar{\lambda}$ is 
\begin{eqnarray}
  \bar{\lambda}^{(2)}&=&\frac{\pi \omega_{p}^2}{8m n_s } \frac{u_\perp^4 c^2}{\Omega^2\gamma_r} \frac{1}{|k_\||}\delta\left(v_\|-\frac{\omega-2\Omega}{k_\|} \right).
  \end{eqnarray}
  The relations between the bounce-averaged coefficients are obtained by the diffusion direction in Eq. (\ref{BtoC}-\ref{BtoF}) and the resonance condition \cite{Harvey:1992},
  \begin{eqnarray}
\langle{\textsf{C} }\rangle_b &=& \langle{\textsf{B} }\rangle_b \frac{1 }{v \sin \vartheta} \left[ \cos \vartheta - \frac{k_{\|} v}{\omega}\right] \frac{\partial \vartheta_0}{ \partial \vartheta}\bigg |_{res} \nonumber \label{BtoC0_B} \\
&=& \langle{\textsf{B} }\rangle_b\frac{\sin \vartheta_0}{v_0 \cos \vartheta_0} \left[ \frac{\cos^2 \vartheta}{\sin^2 \vartheta} - \frac{\omega -m \Omega}{\omega \sin^2 \vartheta}\right]\bigg |_{res}\nonumber \label{BtoC1_B} \\
&=& \langle{\textsf{B} }\rangle_b\frac{(m \Omega_0 /\omega)-\sin^2 \vartheta_0}{v_0 \cos \vartheta_0 \sin \vartheta_0}, \nonumber\\
\langle{\textsf{E} }\rangle_b &=& \langle{\textsf{B} }\rangle_b\frac{(m \Omega_0 /\omega)-\sin^2 \vartheta_0}{v_0 \cos \vartheta_0 }, \nonumber\\
\langle{\textsf{F} }\rangle_b &=& \langle{\textsf{B} }\rangle_b\frac{\{(m \Omega_0 /\omega)-\sin^2 \vartheta_0\}^2}{v_0^2 \cos^2 \vartheta_0 \sin \vartheta_0},  \label{BtoC2_B}
\end{eqnarray}
where the conserved magnetic moment results in ${\partial \vartheta_0}/{ \partial \vartheta}=(\cos\vartheta/\cos\vartheta_0)(\sin\vartheta_0/\sin\vartheta)$, and $\Omega_0=\Omega \sin^2 \vartheta_0/\sin^2 \vartheta$ is the gyrofrequency at the outer-midplane. It is worth noting that the final relations do not depend on the wavevector $k_\|$ and the resonance poloidal locations. Hence, the heavy computation to evaluate the quasilinear tensor is required only in evaluating one component, $\langle{\textsf{B} }\rangle_b $, and other components $\langle{\textsf{C} }\rangle_b $, $\langle{\textsf{E} }\rangle_b $, and $\langle{\textsf{F} }\rangle_b $ are obtained in the post-process using the relations in Eq. (\ref{BtoC2_B}). It is an advantage of using the coordinate in $(u,\vartheta_0)$ that has the invariant variable $u$.

\subsection{Dielectric tensor for non-Maxwellian plasmas in FLR limit}\label{sec:4_2}
The FLR approximation to the quasilinear diffusion coefficients needs to be accordance with the approximation to the imaginary part of dielectric tensor, because we proved $ \langle \mathbf{E}\cdot \mathbf{J} \rangle_w =\dot{W}$ to the lowest order for $n=1$ and $n=2$. For any $n$, the dielectric tensor is generalized in \cite{Stix:AIP1992} by
\begin{eqnarray}
\bar{\chi}_s=\frac{\omega_{p}^2}{\omega} \int_0^{\infty} 2\pi v_\perp dv_\perp \int_{-\infty}^{\infty} dv_\| \left[{\mathbf{e}_\|}{\mathbf{e}_\|}\frac{v_\|^2}{\omega} \left(\frac{1}{v_\|}\frac{\partial f}{\partial v_\|}-\frac{1}{v_\perp}\frac{\partial f}{\partial v_\perp} \right)+\sum_{n=-\infty}^{\infty} \left(\frac{v_\perp U}{\omega-k_\|v_\|-n\Omega}\bar{T}_n\right)\right], 
\end{eqnarray}
where $\bar{T}_n$ is the polarization matrix having the Bessel function $J_n$ and its derivatives $J_n^\prime$. For $n=1$ and $n=2$, the polarization matrix is approximated by the FLR expansion of the Bessel functions  \cite{Phillips2005effects, Bertelli2017full}. For n=1, using $J_1\left({k_\perp v_\perp}/{\Omega} \right)\simeq {k_\perp v_\perp}/{2\Omega}$ and integration by parts for the $v_\perp$ integration, the components of the dielectric tensor in the zero order are \cite{Phillips2005effects}
\begin{eqnarray}
\chi_{xx}^{(0)}=\chi_{yy}^{(0)}=\frac{\omega_{ps}^2}{\omega} \left[\frac{1}{2} \left( \tilde{A}_{1,0} +\tilde{A}_{-1,0} \right)\right] ,\nonumber\\
\chi_{xy}^{(0)}=-\chi_{yx}^{(0)}=i\frac{\omega_{ps}^2}{\omega} \left[\frac{1}{2} \left( \tilde{A}_{1,0} -\tilde{A}_{-1,0} \right)\right], \label{chi0}
\end{eqnarray}
where $\tilde{A}_{n,j}$ has the integration in $v_\|$,
\begin{eqnarray}
\tilde{A}_{n,j}&=& \int_{-\infty}^{\infty} dv_\|  \frac{1}{\omega-k_\|v_\|-n\Omega}  \int_0^{\infty} 2\pi v_\perp dv_\perp H_j(v_\|,v_\perp),\\
H_0(v_\|,v_\perp)&=&\frac{1}{2}\frac{k_\|w_\perp^2}{\omega}\frac{\partial f}{\partial v_\|}-\left(1-\frac{k_\|v_\|}{\omega}\right)f.\label{H0_chi}
\end{eqnarray}
Here, $w_\perp$ is the average perpendicular velocity. This approximation is exactly corresponding to the quasilinear diffusion approximation in $\dot{W}^{n=1(0)}$ of Eq. (\ref{WFLR0_n1}), because the operator $H_0$ in Eq. (\ref{H0_chi}) corresponds to the operator $U$ in Eq. (\ref{U_dql}) that determines the diffusion direction, and the left-hand polarization in Eq. (\ref{chi0}) corresponds to the dielectric constant for $E_+$. Thus, for the dielectric tensor of the non-Maxweliian distribution, the operator $\tilde{A}_{1,0}$ can be used instead of $(\hat{L}+\Delta\hat{L_1})$ in TORIC. 

For $n=2$, the imaginary part is determined by
\begin{eqnarray}
\chi_{xx}^{n=2,(2)}=\chi_{yy}^{n=2,(2)}=\frac{\omega_{ps}^2}{\omega} \left[\frac{k_\perp^2w_\perp^2}{4\Omega^2} \left( \tilde{A}_{2,1} +\tilde{A}_{-2,1} \right)\right], \nonumber\\
\chi_{xy}^{n=2,(2)}=-\chi_{yx}^{n=2,(2)}=i\frac{\omega_{ps}^2}{\omega} \left[\frac{k_\perp^2w_\perp^2}{4\Omega^2} \left( \tilde{A}_{2,1} -\tilde{A}_{-2,1} \right)\right], \label{chi_n2}
\end{eqnarray}
where $J_2 \left({k_\perp v_\perp}/{\Omega} \right)\simeq {k_\perp^2 v_\perp^2}/{8\Omega^2}$ is used to the lowest order  \cite{Phillips2005effects}, giving
\begin{eqnarray}
H_1(v_\|,v_\perp)&=&\frac{1}{4}\frac{k_\|w_\perp^2}{\omega}\frac{\partial f}{\partial v_\|}\frac{v_\perp^4}{w_\perp^4}-\left(1-\frac{k_\|v_\|}{\omega}\right)f\frac{v_\perp^2}{w_\perp^2}.
\end{eqnarray}
Here, $k_\perp$ in Eq. (\ref{chi_n2}) is not explicitly evaluated in TORIC but the corresponding vector operators in Eq. (\ref{J2_2_toric}) are used. Then, the operator $\tilde{A}_{2,1}$ can be used instead of $\hat{\lambda}^{(2)} + \Delta \hat{\lambda}^{(2)}_1$ in Eq. (\ref{J2_2_toric}), giving the $ \langle \mathbf{E}\cdot \mathbf{J} \rangle_w =\dot{W}$ to the lowest order of $n=2$ in TORIC. 
\clearpage

\section{Examples}\label{sec:5}
We present some examples using the reduced model in TORIC-CQL3D and compare them with the results by the full model in AORSA-CQL3D \cite{jaeger2006self}. In the following two examples, we simulate the 10MW ICRF minority species heating scenarios in ITER with a static magnetic field 5.3T at the magnetic axis, as in the benchmark study of \cite{budny2012benchmarking}. The two examples have the different wave frequencies and the cyclotron layer location is set to be off-axis in the first example and on-axis in the second example. The wave power density of the first example is much smaller than the second examples because the off-axis damping of the first example results in the wave energy transfer at a larger volume compared to the core damping of the second example. Since the maximum energy of the fast ions depends on the wave power density, we can compare the validity of the FLR approximation in the examples.

\subsection{Fundamental damping by He3 with 48 MHz ICRF in ITER}\label{sec:5_1}
In this example, we simulate three ion species with the ratio of (D,T,He3)$=(48,48,2)\%$ and the ICRF wave frequency is 48MHz. The dominant wave power is absorbed by the minority species He3 in the off-axis because the cyclotron layer is located at $R=R_0+0.87 \textrm{m}$ in the low field side, which is tangential to the flux surface of $r/a=0.52$. Here, $R_0=6.2 \textrm{m}$ is the major radius of magnetic-axis and $r/a$ is the normalized radial coordinate, which is determined by the square root of poloidal flux in this section. The power absorption results between TORIC and AORSA are similar particularly in the ion damping, but some differences are seen in the electron channel due to the approximation made in TORIC for ion Bernstein wave damping \cite{brambilla1999numerical}. The power decomposition is $67\%$ of He3 fundamental damping, $17\%$ of T second harmonic damping, and 15\% electron damping in TORIC, while it is $79\%$ of He3 fundamental damping, $13\%$ of T second harmonic damping, and $9\%$ electron damping in AORSA. The radial power profiles of He3 are reasonably similar between two codes as shown in Figure 1. 

The wave power density ($\lesssim 1\textrm{MW/m}^3$) in this example results in the non-negligible change from the initial Maxwellian distribution as shown in Figure 2, although its impact on the power absorption profile is not significant as shown in the difference between blue and green curves in Figure 1. The patterns of the distribution functions show the reasonable agreement between the reduced model (TORIC-CQL3D) in Figure 2-(a) and (c) and the full model (AORSA-CQL3D) in Figure 2-(b) and (d). Figure 3-(a) and (b) shows the diagonal component of the quasilinear diffusion coefficient in the speed direction for the reduced model and the full model. Their patterns are reasonably similar, while the diffusion of the high energetic ions around 1MeV in the reduced model is higher than that of the full model. In Figure 3-(a), the diffusion coefficients in the reduced model almost constantly increase in terms of perpendicular velocity, which is relevant to $v_\perp^2$ in Eq. (\ref{Bn_1}) that does not have the decaying factor by the Bessel function $J_0(k_\perp \rho_i)^2$ in Eq. (\ref{Bfull}). Nevertheless, this difference does not affect significantly on the energy transfer or the distribution functions in Figure 2. The figures of Figure 3-(c) and (d) show the small difference of diffusion coefficients between two models if they are multiplied by the weight of the general exponential decay of the distribution function. In this example, the maximum energy of the energetic ions is approximately 1MeV with a Larmor radius of about 3 cm and the fast wave branch has approximately 20 cm wavelength. Accordingly, the most of ions satisfy $k_\perp \rho_i \lesssim 1$ and the reduced model is marginally valid.   
 
\begin{figure*}
(a) \;\;\;\;\;\;\;\;\;\;\;\;\;\;\;\;\;\;\;\;\;\;\;\;\;\;\;\;\;\;\;\;\;\;\;\;\;\;\;\;\;\;\;\;\;\;\;\;\;\;\;\;\;\;\;\;\;\;\;\;\;\;\;\;\;\;\;\;\;\;\;\;(b) \;\;\;\;\;\;\;\;\;\;\;\;\;\;\;\;\;\;\;\;\;\;\;\;\;\;\;\;\;\;\;\;\;\;\;\;\;\;\;\;\;\;\;\;\;\;\;\;\;\;\;\;\;\;\;\;\;\;\;\;\;\;\;\\
\includegraphics[scale=0.4]{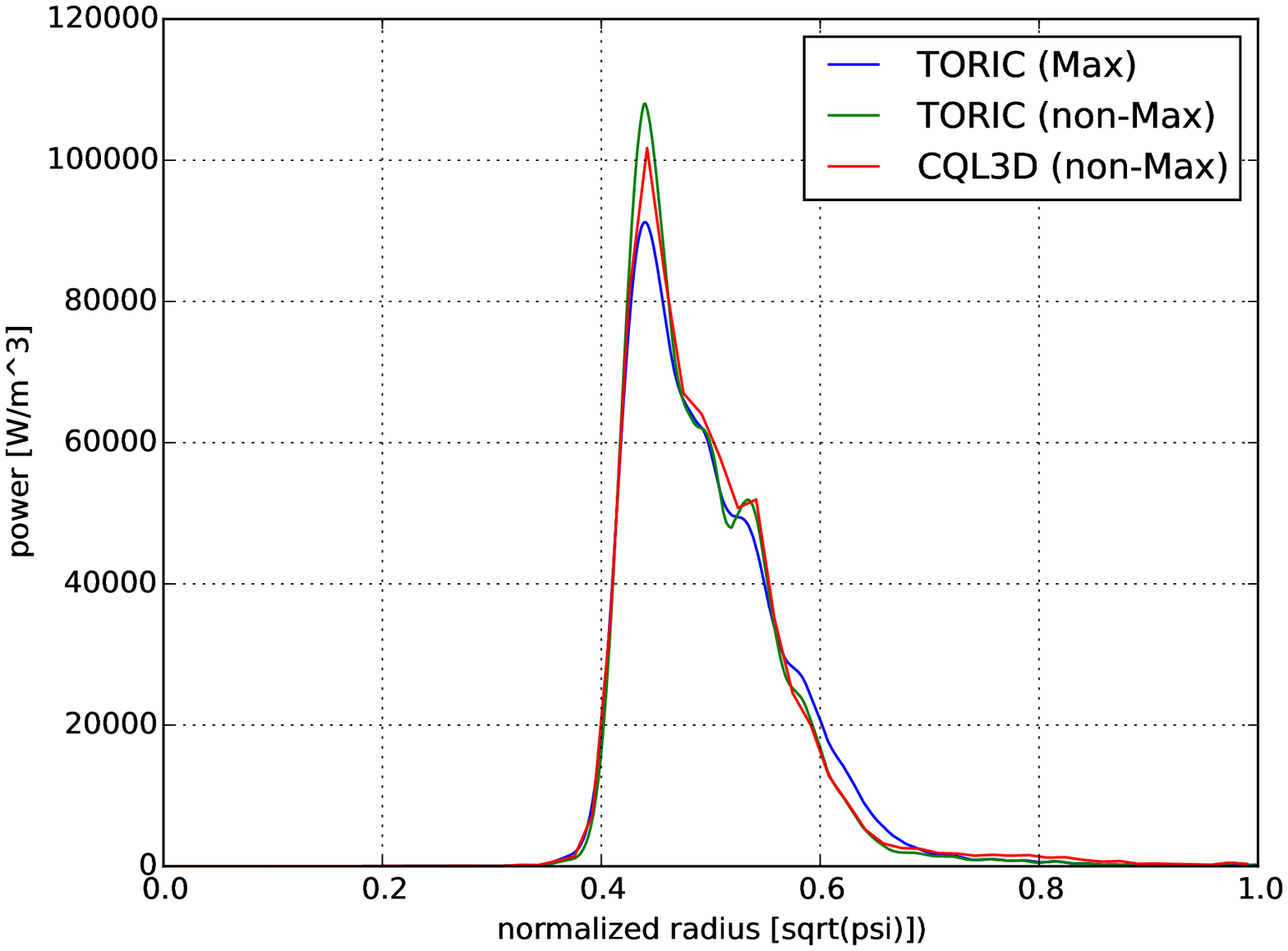}\includegraphics[scale=0.4]{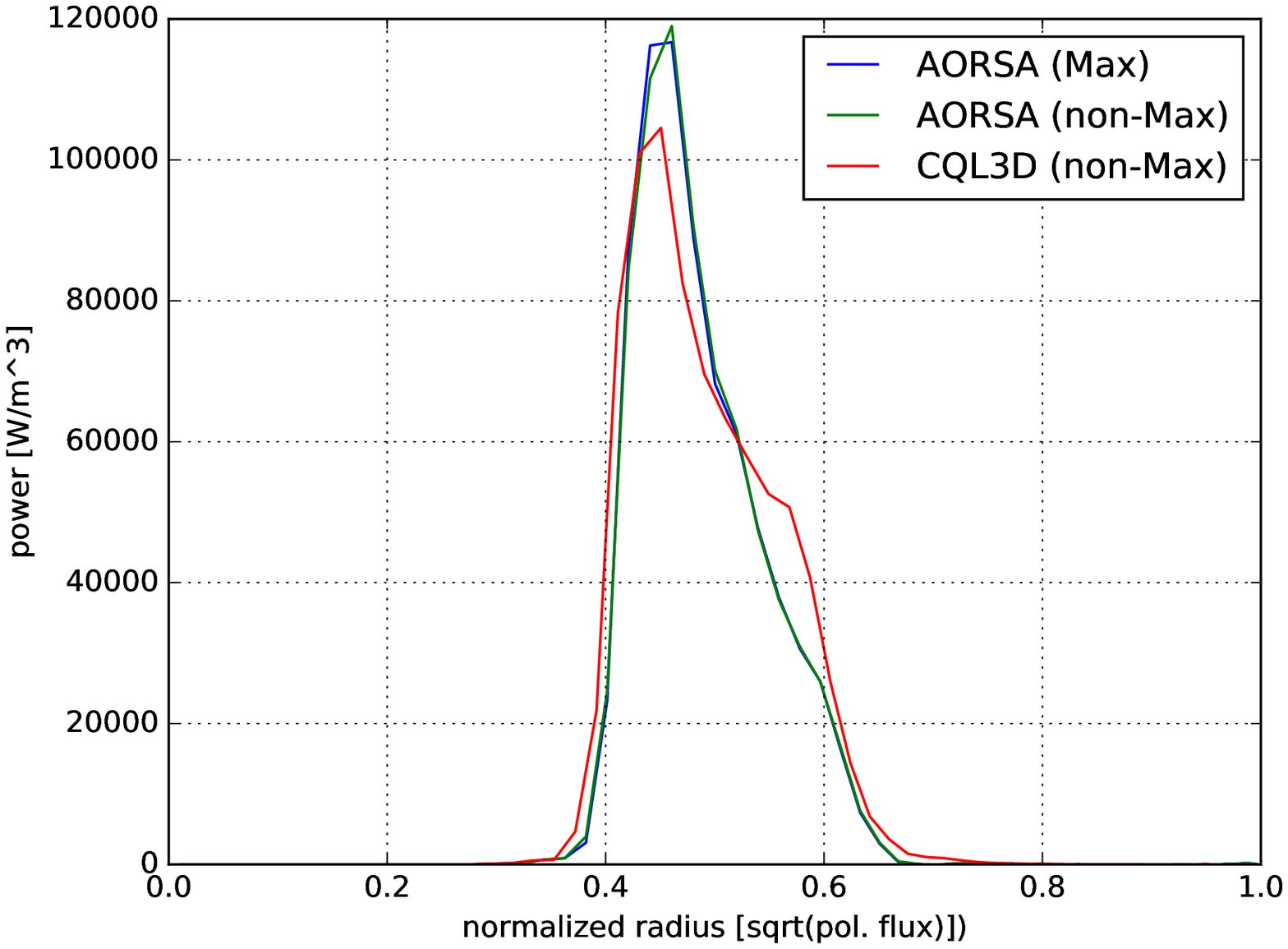}
\caption{Example 1: radial profiles of power absorption by He3 for the 48 MHz ICRF injection. The total wave power is 10MW, and the profiles are simulated (a) by the reduced model in TORIC-CQL3D and (b) by the full model in AORSA-CQL3D. The blue curve is the absorption by $ \langle \mathbf{E}\cdot \mathbf{J} \rangle_w$ for the isotropic Maxwellian distribution function. The green and red curves are the absorption by $ \langle \mathbf{E}\cdot \mathbf{J} \rangle_w$ in the wave code (TORIC or AORSA) and the absorption by $\dot{W}$ in the Fokker-Plank code (CQL3D), respectively, when the iteration converges so that the self-consistent non-Maxwellian distribution is developed.} 
\end{figure*}

\begin{figure*}
(a) \;\;\;\;\;\;\;\;\;\;\;\;\;\;\;\;\;\;\;\;\;\;\;\;\;\;\;\;\;\;\;\;\;\;\;\;\;\;\;\;\;\;\;\;\;\;\;\;\;\;\;\;\;\;\;\;\;\;\;\;\;\;\;\;\;\;\;\;\;\;\;\;(b) \;\;\;\;\;\;\;\;\;\;\;\;\;\;\;\;\;\;\;\;\;\;\;\;\;\;\;\;\;\;\;\;\;\;\;\;\;\;\;\;\;\;\;\;\;\;\;\;\;\;\;\;\;\;\;\;\;\;\;\;\;\;\;\\
\includegraphics[scale=0.4]{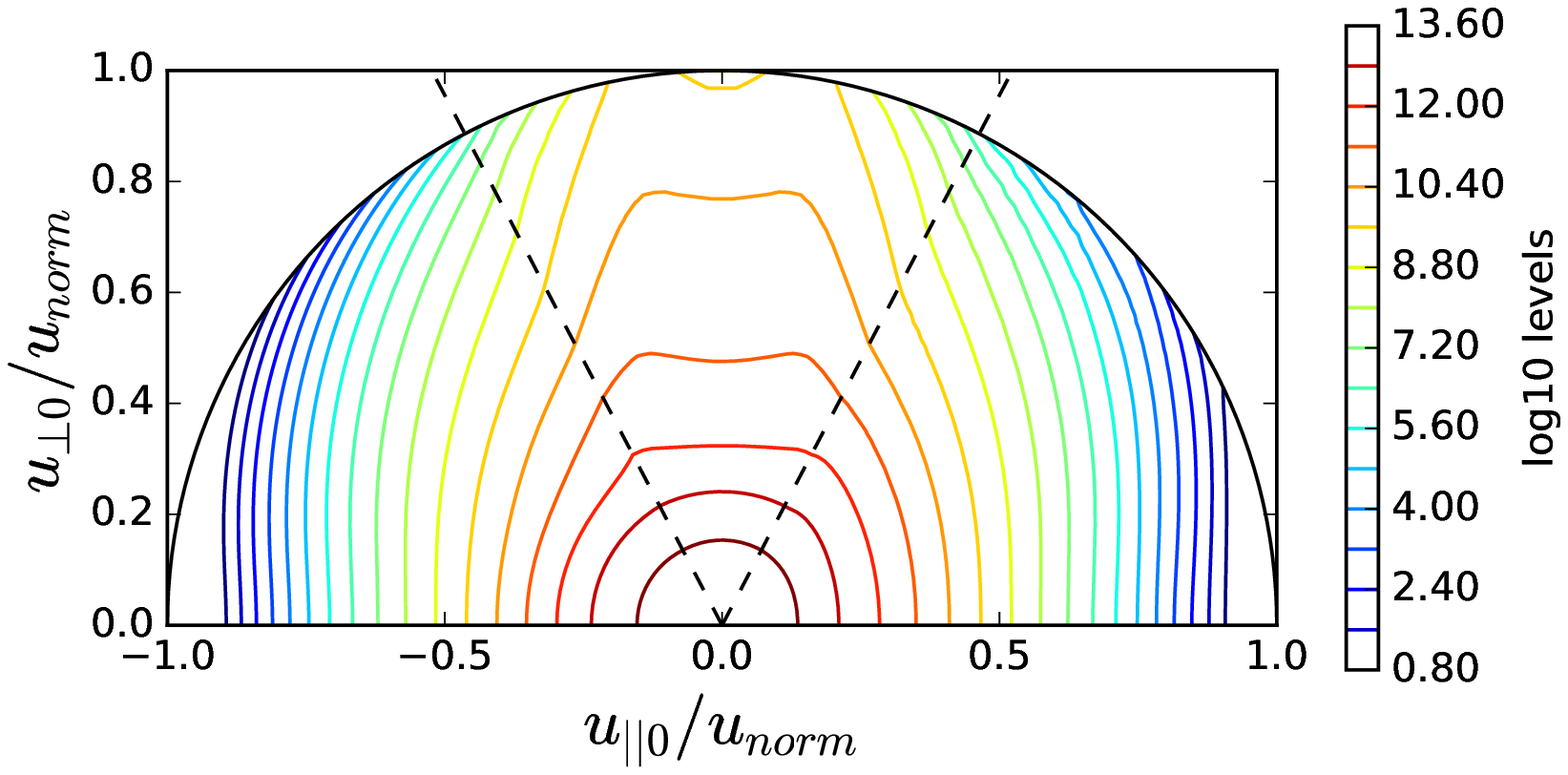}\includegraphics[scale=0.4]{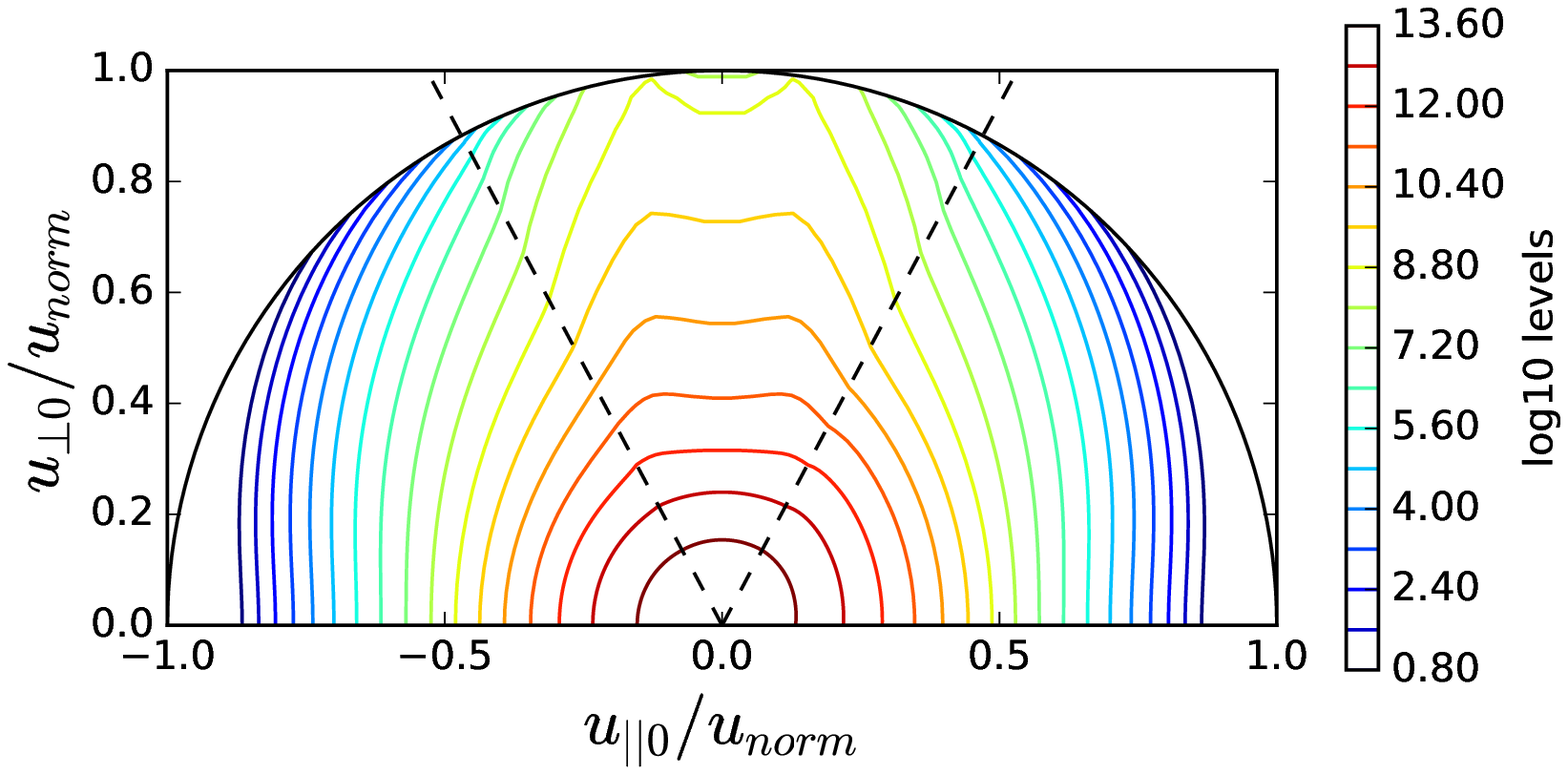}\\
(c) \;\;\;\;\;\;\;\;\;\;\;\;\;\;\;\;\;\;\;\;\;\;\;\;\;\;\;\;\;\;\;\;\;\;\;\;\;\;\;\;\;\;\;\;\;\;\;\;\;\;\;\;\;\;\;\;\;\;\;\;\;\;\;\;\;\;\;\;\;\;\;\;(d) \;\;\;\;\;\;\;\;\;\;\;\;\;\;\;\;\;\;\;\;\;\;\;\;\;\;\;\;\;\;\;\;\;\;\;\;\;\;\;\;\;\;\;\;\;\;\;\;\;\;\;\;\;\;\;\;\;\;\;\;\;\;\;\\
\includegraphics[scale=0.3]{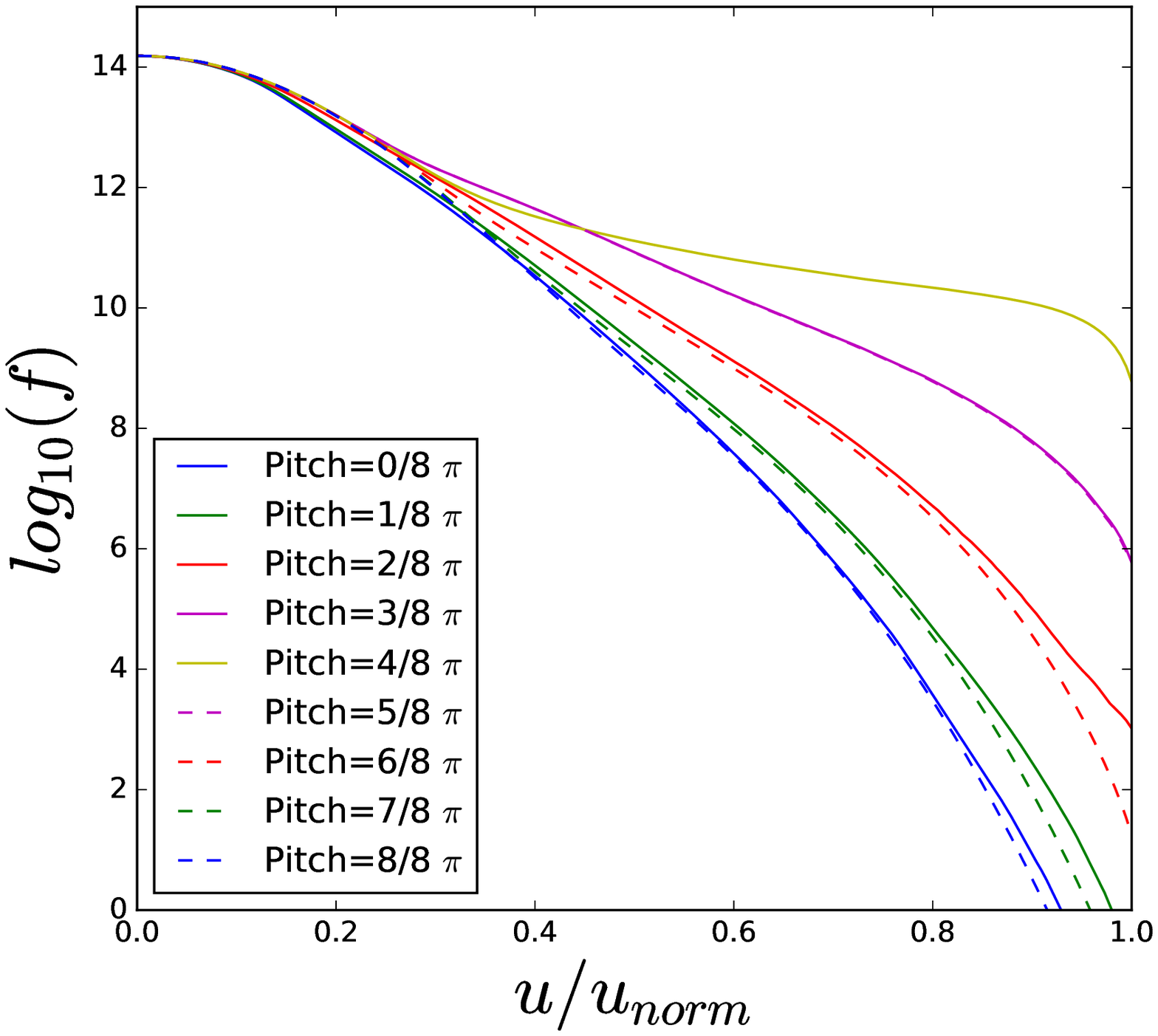}\;\;\;\;\;\;\;\;\;\;\;\;\includegraphics[scale=0.3]{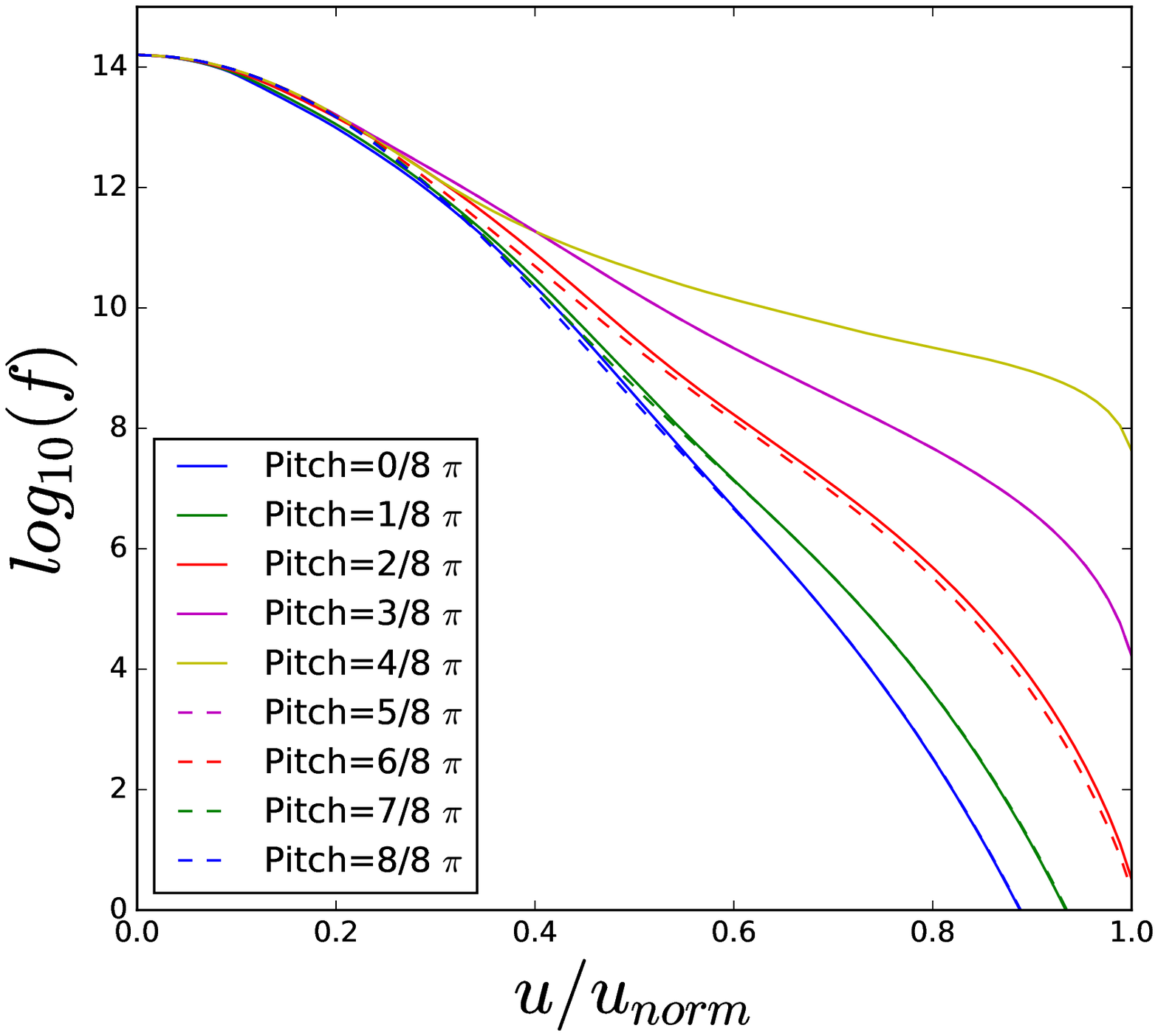}
\caption{Example 1: (a) and (b) are the 2-D contour plots of the distribution function in $(u_{\|0},u_{\perp0})$. (c) and (d) are the 1-D distribution functions in terms of $u$ for several pitch-angles $\vartheta$ at $r/a=0.44$. Figure (a) and (c) are simulated by TORIC-CQL3D and (b) and (d) are simulated by AORSA-CQL3D, where $u_{norm}$ is the momentum corresponding to the energy of He3 1MeV. The dashed lines in the contour plots are the trapped-passing boundaries, and the unit of the distribution function is $\textrm{cm}^{-3}$.}
\end{figure*}

\begin{figure*}
(a) \;\;\;\;\;\;\;\;\;\;\;\;\;\;\;\;\;\;\;\;\;\;\;\;\;\;\;\;\;\;\;\;\;\;\;\;\;\;\;\;\;\;\;\;\;\;\;\;\;\;\;\;\;\;\;\;\;\;\;\;\;\;\;\;\;\;\;\;\;\;\;\;(b) \;\;\;\;\;\;\;\;\;\;\;\;\;\;\;\;\;\;\;\;\;\;\;\;\;\;\;\;\;\;\;\;\;\;\;\;\;\;\;\;\;\;\;\;\;\;\;\;\;\;\;\;\;\;\;\;\;\;\;\;\;\;\;\\
\includegraphics[scale=0.20]{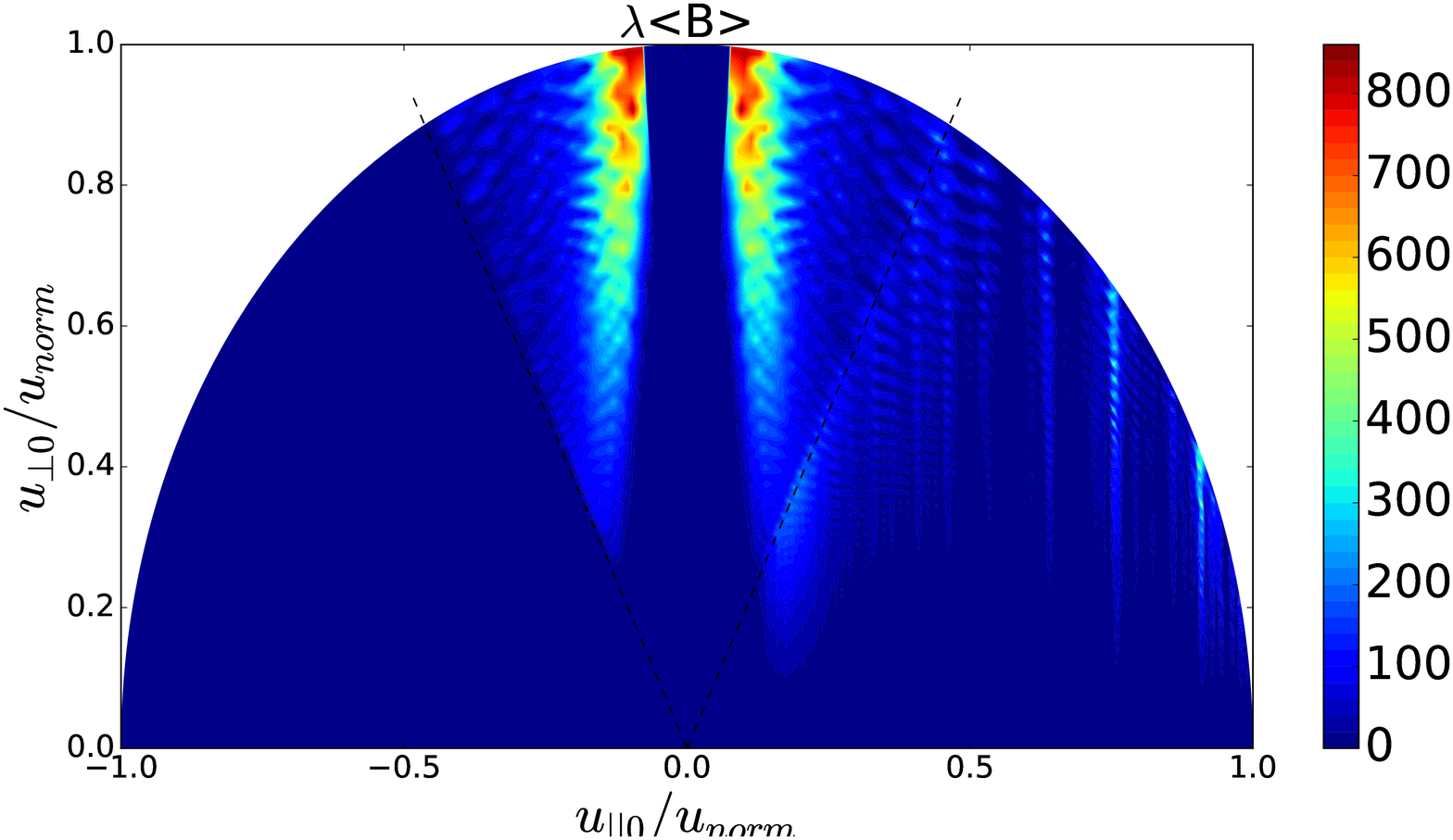}\includegraphics[scale=0.20]{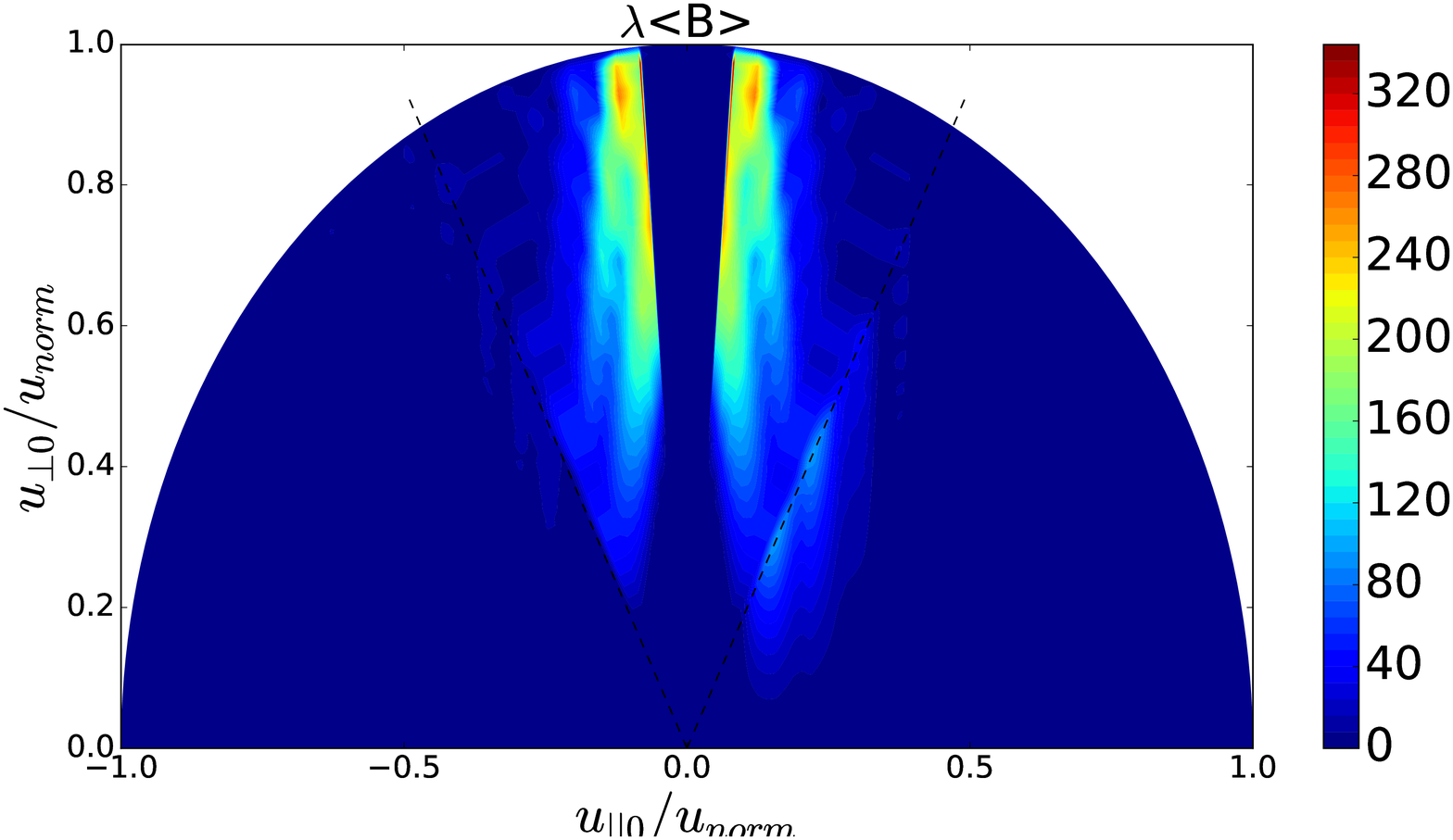}\\
(c) \;\;\;\;\;\;\;\;\;\;\;\;\;\;\;\;\;\;\;\;\;\;\;\;\;\;\;\;\;\;\;\;\;\;\;\;\;\;\;\;\;\;\;\;\;\;\;\;\;\;\;\;\;\;\;\;\;\;\;\;\;\;\;\;\;\;\;\;\;\;\;\;(d) \;\;\;\;\;\;\;\;\;\;\;\;\;\;\;\;\;\;\;\;\;\;\;\;\;\;\;\;\;\;\;\;\;\;\;\;\;\;\;\;\;\;\;\;\;\;\;\;\;\;\;\;\;\;\;\;\;\;\;\;\;\;\;\\
\includegraphics[scale=0.20]{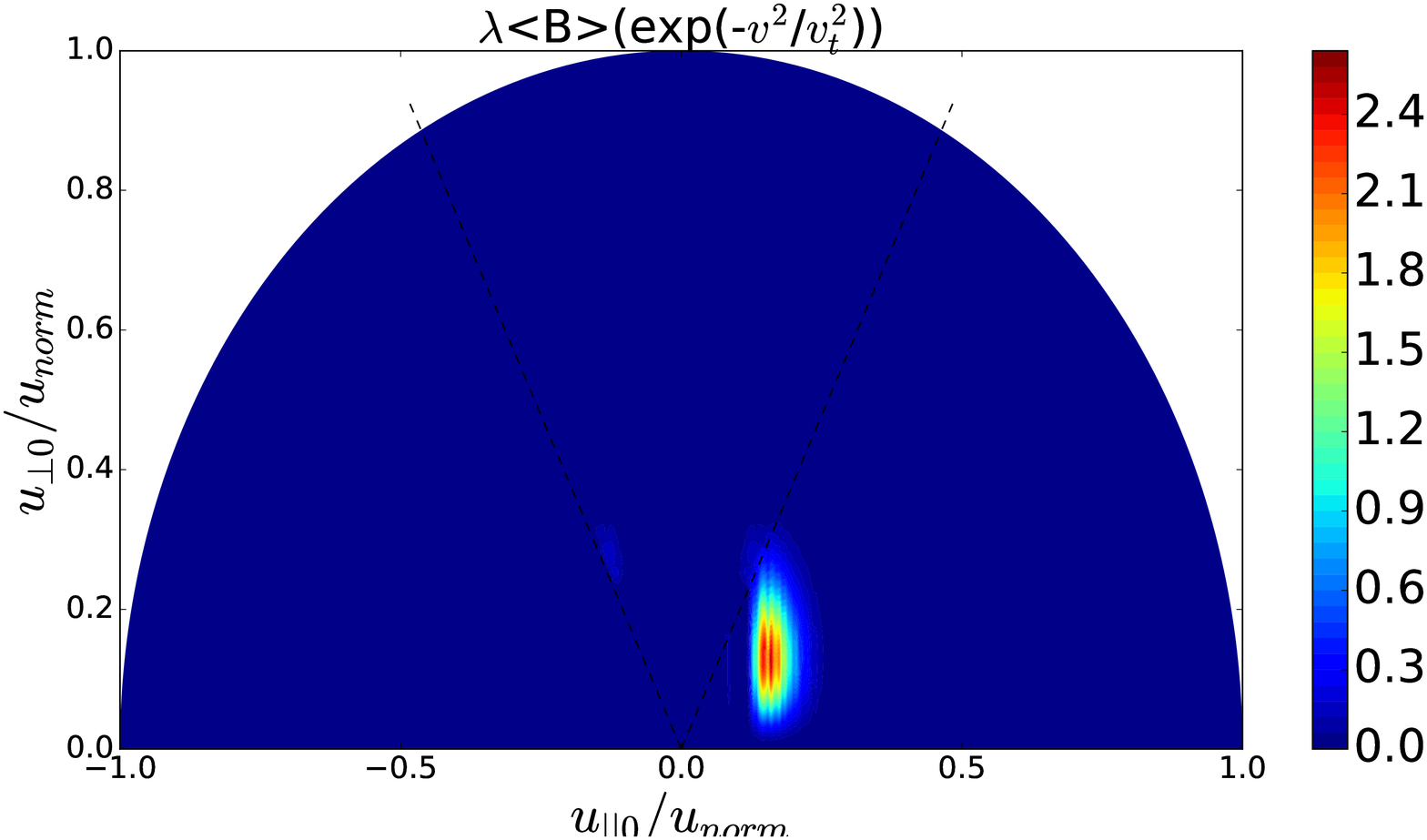}\includegraphics[scale=0.20]{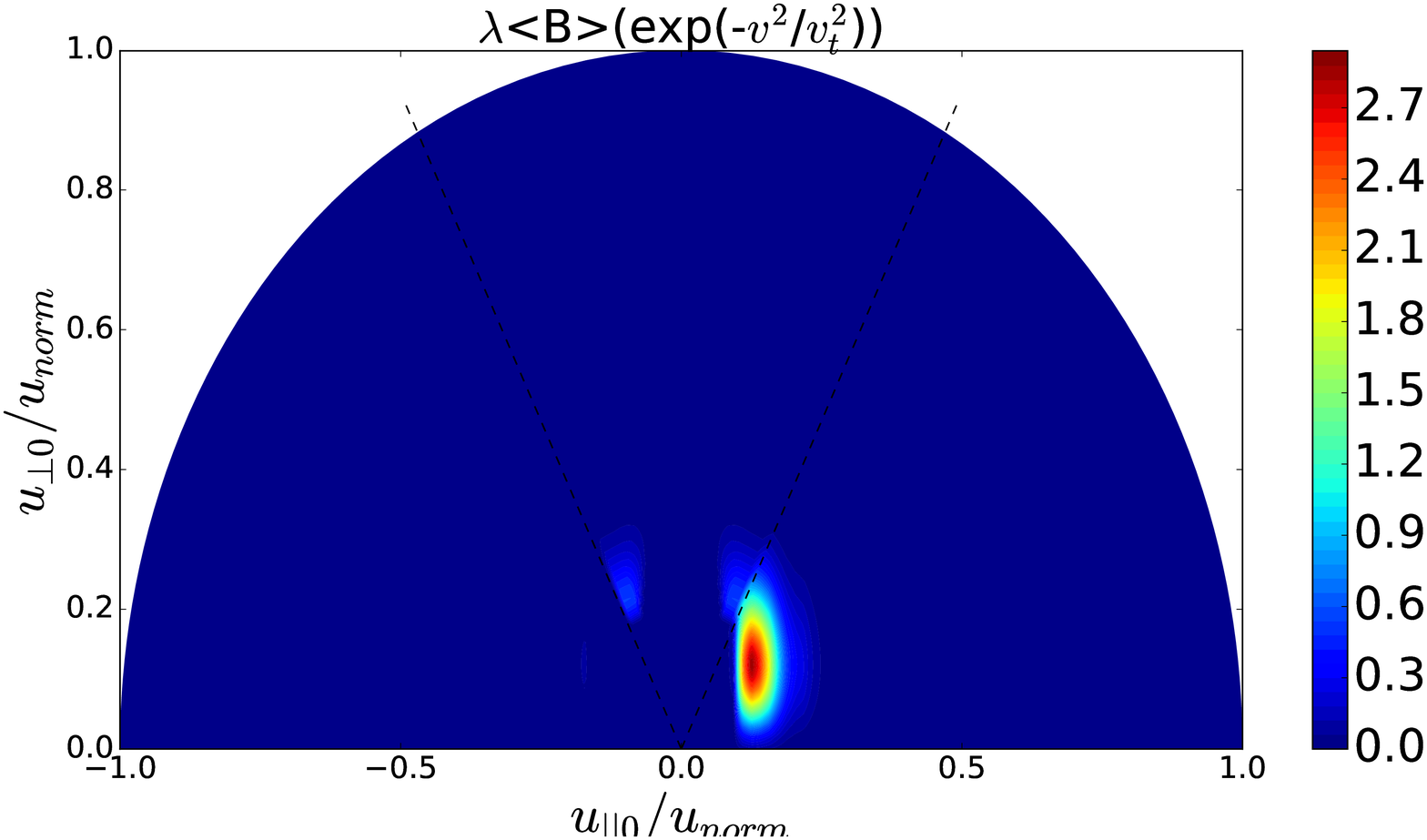}
\caption{Example 1: (a) and (b) are the 2-D contour plots of quasilinear diffusion coefficient $\lambda \langle B\rangle$ in $(u_{\|0},u_{\perp0})$ and (c) and (d) are those weighted by the Maxwellian factor $\lambda \langle B\rangle \exp(-v^2/v_t^2)$ at $r/a=0.44$. Figure (a) and (c) are simulated by TORIC-CQL3D and (b) and (d) are simulated by AORSA-CQL3D, where the contour values are normalized to $u_{norm}$ is the momentum corresponding to the energy of He3 1MeV. The dashed lines are the trapped-passing boundaries, and the unit of the  $\lambda \langle B\rangle$ is $v_{norm}^{4}$ where $v_{norm}$ is the speed corresponding to $u_{norm}$.} 
\end{figure*}

\subsection{Fundamental damping by He3 with 52.5 MHz ICRF in ITER}\label{sec:5_2}
In this example, we simulate three ion species with the ratio of (D,T,He3)$=(48,48,2)\%$ as the first example, but the ICRF wave frequency is 52.5MHz. This example has the same condition as the benchmark case between TORIC and AORSA in \cite{budny2012benchmarking}. The dominant wave power is absorbed by the minority species He3 around the magnetic axis because the cyclotron layer is located at $R=R_0+0.17 m$ in the low field side, which is tangential to the flux surface of $r/a=0.10$. The power decompositions reasonably agree between TORIC and AORSA: For Maxwellian plasmas, $51\%$ of He3 fundamental damping, $15\%$ of T second harmonic damping, and $34\%$ electron damping in TORIC, while $55\%$ of He3 fundamental damping, $12\%$ of T second harmonic damping, and $34\%$ electron damping in AORSA.  For self-consistent non-Maxwellian plasmas, we found $50\%$ of He3 fundamental damping, $16\%$ of T second harmonic damping, and $34\%$ electron damping in TORIC, while $56\%$ of He3 fundamental damping, $12\%$ of T second harmonic damping, and $33\%$ electron damping in AORSA. These results also agree with the previous results in \cite{budny2012benchmarking}. Figure 4 shows the radial power profiles of He3 for this example. Both in TORIC and AORSA, the damping at the core ($r/a\lesssim 0.1$) is reduced by the non-Maxwellian plasmas and the power absorption profile is broadened (compare the blue curve and the red curve). The broaden profile of the power absorption with the non-Maxwellian distribution is expected because of the Doppler shift of the energetic ions \cite{dumont2005effects}. 

The difference between the green curve and the red curve in TORIC-CQL3D of Figure 4-(a) indicates some problems in the iteration convergence. The high power density ($>5\textrm{MW/m}^3$) at the core results in the high energetic tail up to 4 MeV in Figure 5, in which the FLR approximation may not be acceptable. For the high energetic ions with $k_\perp \rho_i \geq 2$, their quasilinear diffusion coefficients in the reduced model of Eq. (\ref{Bn_1}) may be inaccurate compared to the full model of Eq. (\ref{Bfull}) due to two reasons. One reason is the missing Bessel function factor $J_0^2 (k_\perp \rho_i \geq 2)<0.2$ in the term of $E_+$, and the other reason is the missing higher order term of $J_2(k_\perp \rho_i) E_-$. The former likely causes the overestimation of the diffusion, while the latter causes the underestimation when $J_2$ is not negligible for the high $k_\perp \rho_i$. Figure 6-(a) and (b) show such differences, in which the diffusion of the reduced model increases in $v_\perp$ while the diffusion of the full model is small up to the particle energy 2MeV but it is large beyond the energy. 

For the high energetic particles ($>2\textrm{MeV}$), the distribution function of the full model in Figure 5-(b) and (d) is much larger than that of the reduced model in Figure 5-(a) and (c) because of the strong diffusion. Nevertheless, for the most population particles below 2MeV the distribution functions in Figure 5 are very similar between two models because the diffusions are comparable according to the Figure 6-(c) and (d). Thus, even for the simulation of the high wave power density which results in the problematic FLR approximation, the reduced model can be useful to estimate the sub-MeV distribution functions and the wave power absorption.   
\begin{figure*}
(a) \;\;\;\;\;\;\;\;\;\;\;\;\;\;\;\;\;\;\;\;\;\;\;\;\;\;\;\;\;\;\;\;\;\;\;\;\;\;\;\;\;\;\;\;\;\;\;\;\;\;\;\;\;\;\;\;\;\;\;\;\;\;\;\;\;\;\;\;\;\;\;\;(b) \;\;\;\;\;\;\;\;\;\;\;\;\;\;\;\;\;\;\;\;\;\;\;\;\;\;\;\;\;\;\;\;\;\;\;\;\;\;\;\;\;\;\;\;\;\;\;\;\;\;\;\;\;\;\;\;\;\;\;\;\;\;\;\\
\includegraphics[scale=0.4]{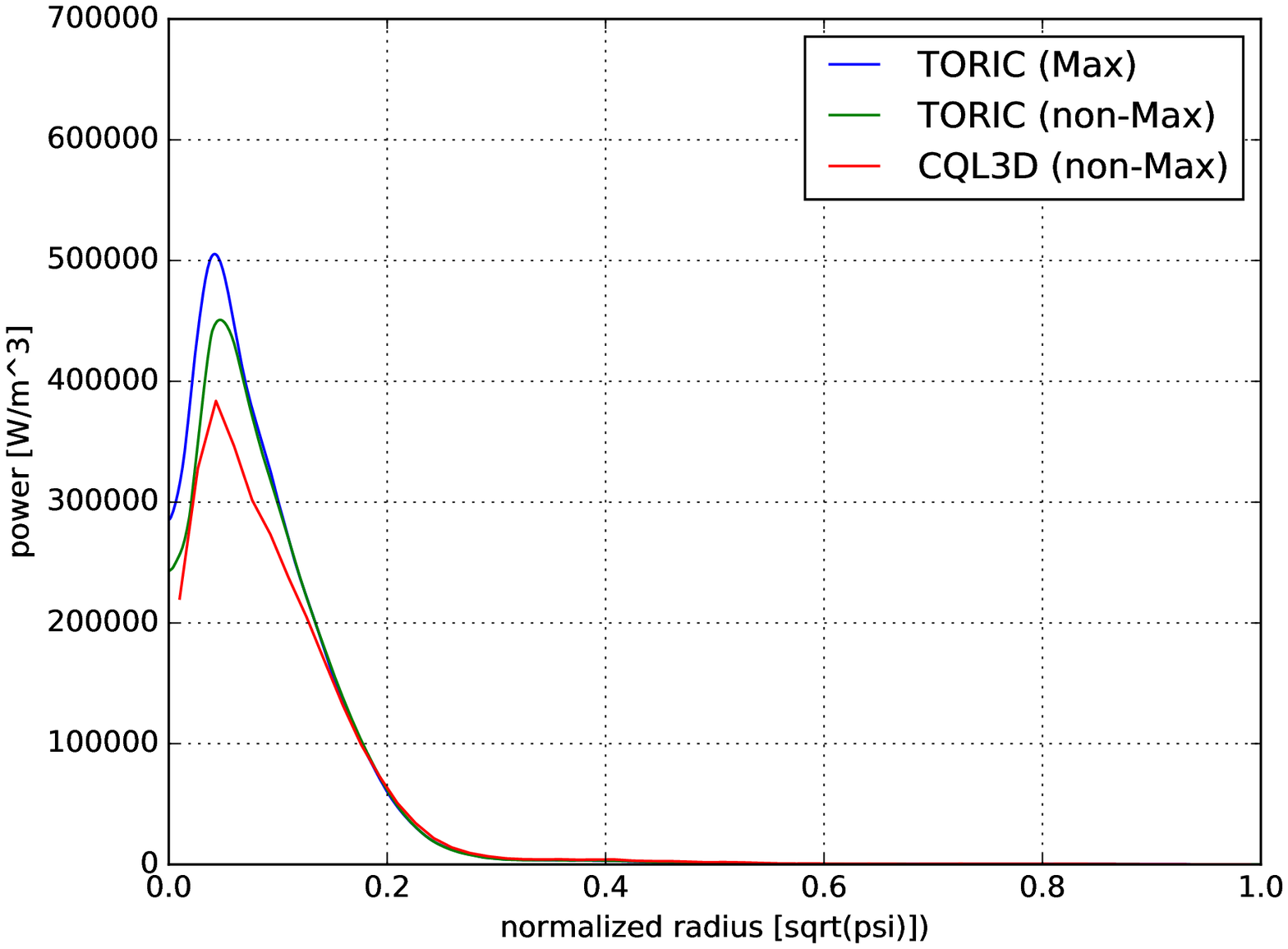}\includegraphics[scale=0.4]{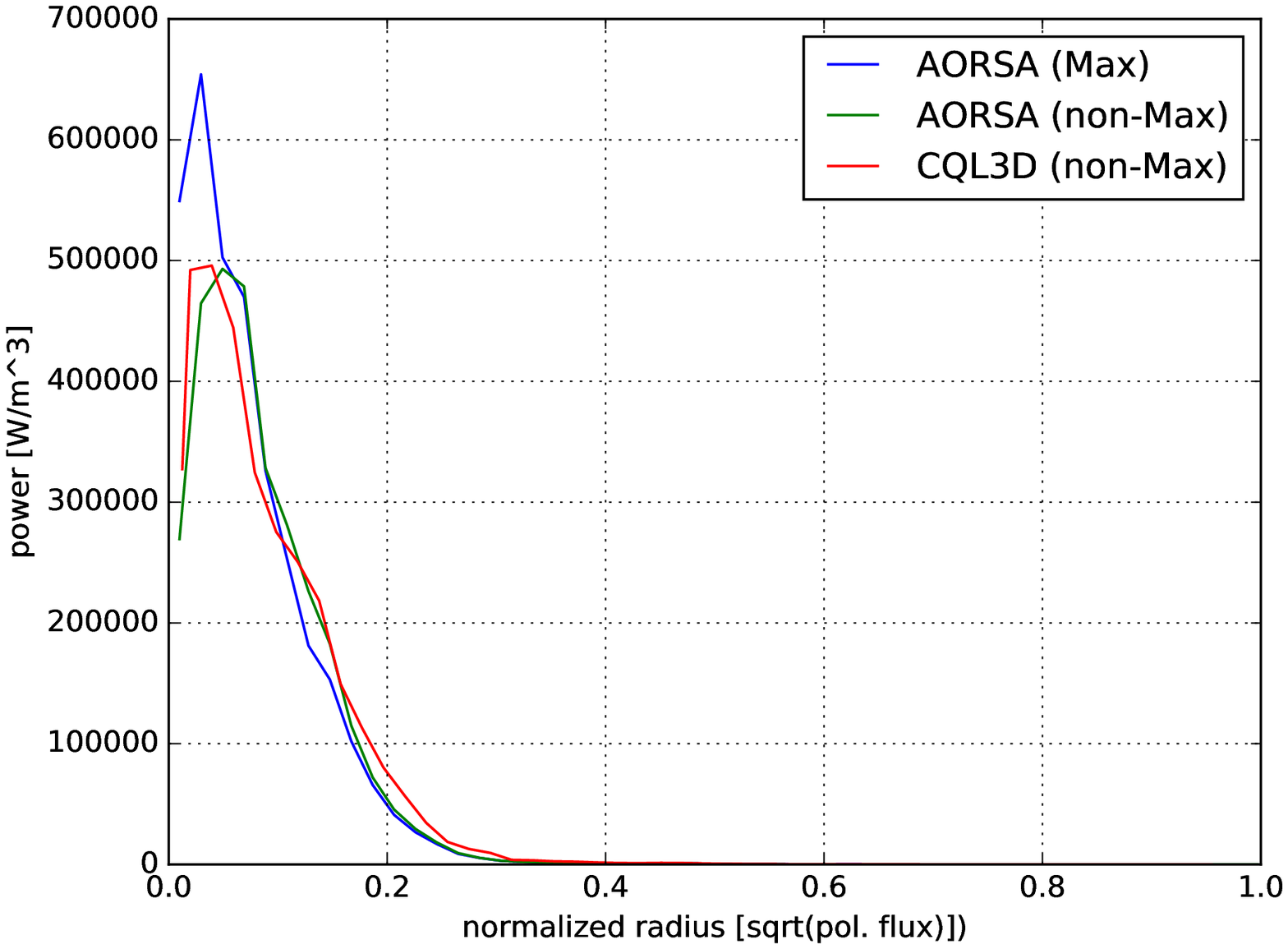}
\caption{Example 2:  radial profiles of power absorption by He3 for the 52.5 MHz ICRF injection. The total wave power is 10MW, and the profiles are simulated (a) by the reduced model in TORIC-CQL3D and (b) by the full model in AORSA-CQL3D.}
\end{figure*}

\begin{figure*}
(a) \;\;\;\;\;\;\;\;\;\;\;\;\;\;\;\;\;\;\;\;\;\;\;\;\;\;\;\;\;\;\;\;\;\;\;\;\;\;\;\;\;\;\;\;\;\;\;\;\;\;\;\;\;\;\;\;\;\;\;\;\;\;\;\;\;\;\;\;\;\;\;\;(b) \;\;\;\;\;\;\;\;\;\;\;\;\;\;\;\;\;\;\;\;\;\;\;\;\;\;\;\;\;\;\;\;\;\;\;\;\;\;\;\;\;\;\;\;\;\;\;\;\;\;\;\;\;\;\;\;\;\;\;\;\;\;\;\\
\includegraphics[scale=0.4]{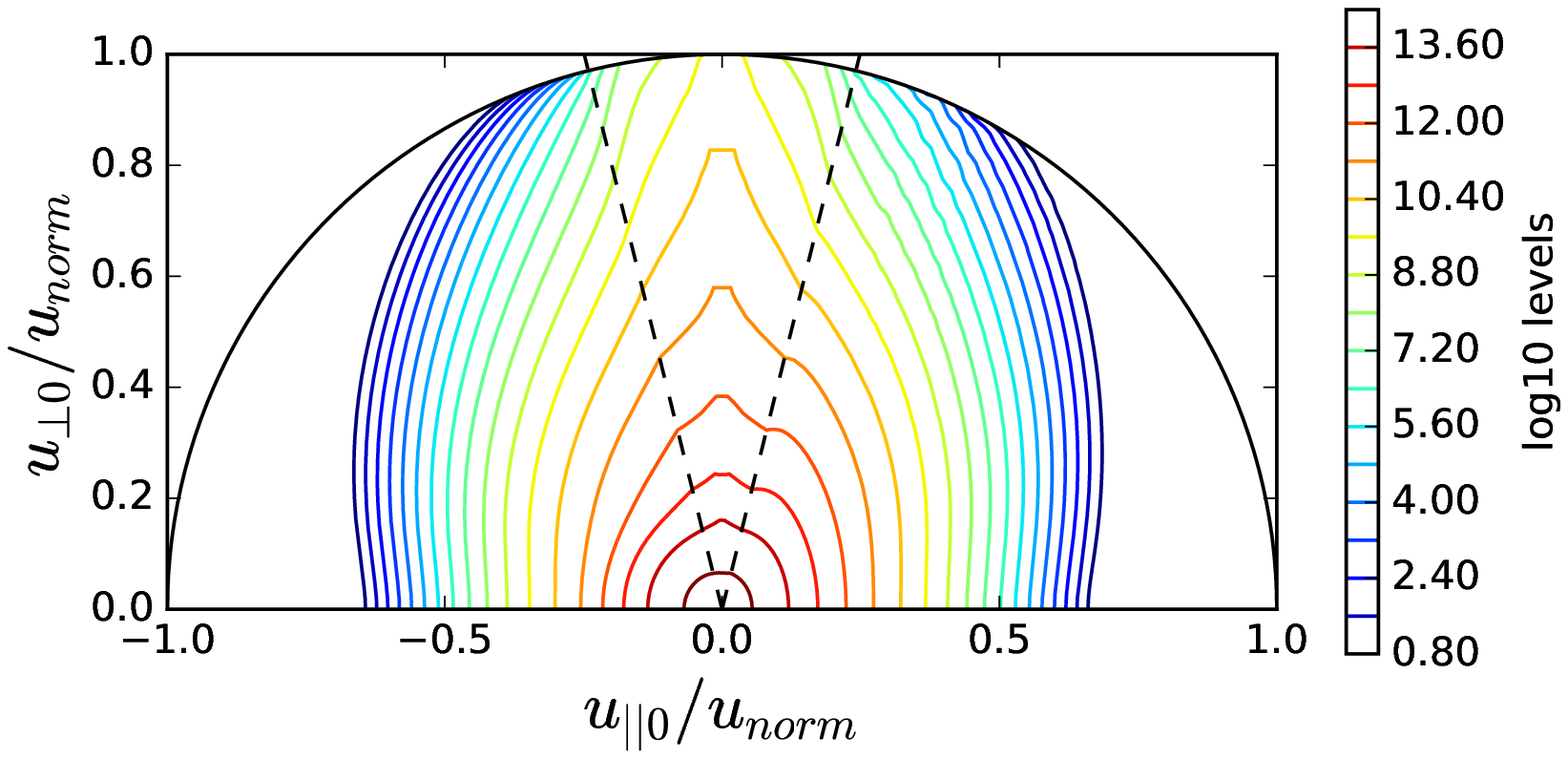}\includegraphics[scale=0.4]{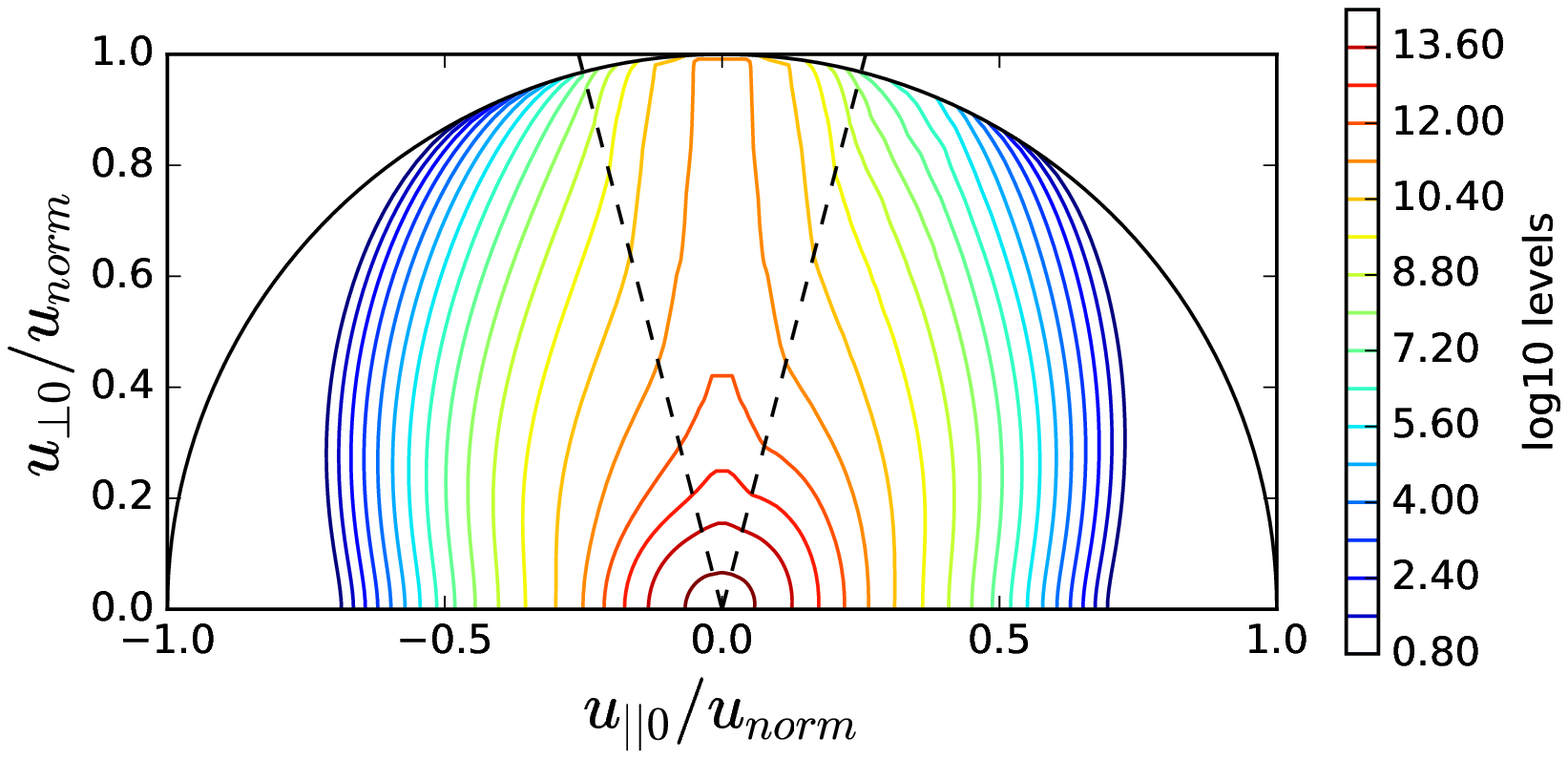}\\
(c) \;\;\;\;\;\;\;\;\;\;\;\;\;\;\;\;\;\;\;\;\;\;\;\;\;\;\;\;\;\;\;\;\;\;\;\;\;\;\;\;\;\;\;\;\;\;\;\;\;\;\;\;\;\;\;\;\;\;\;\;\;\;\;\;\;\;\;\;\;\;\;\;(d) \;\;\;\;\;\;\;\;\;\;\;\;\;\;\;\;\;\;\;\;\;\;\;\;\;\;\;\;\;\;\;\;\;\;\;\;\;\;\;\;\;\;\;\;\;\;\;\;\;\;\;\;\;\;\;\;\;\;\;\;\;\;\;\\
\includegraphics[scale=0.3]{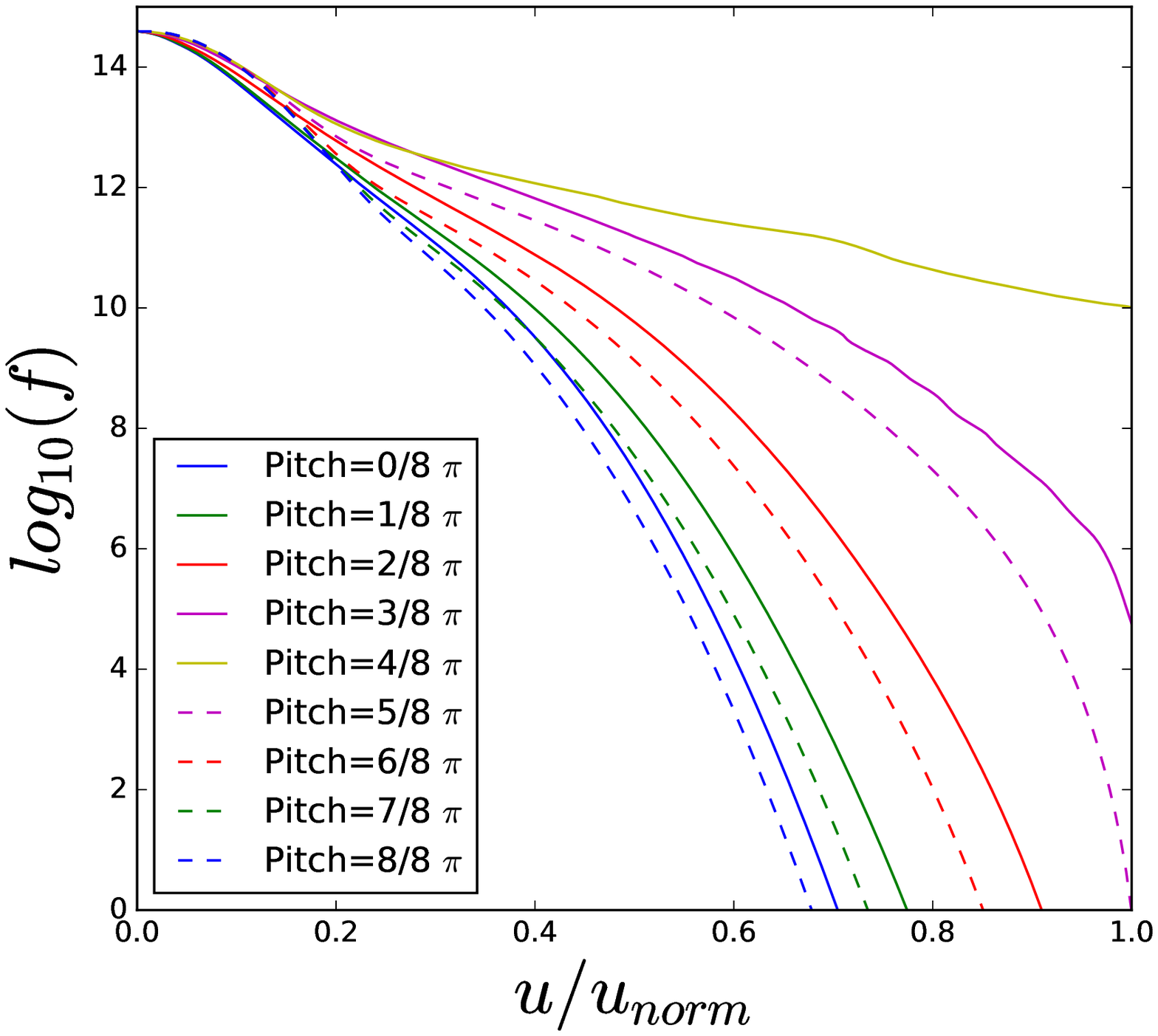}\;\;\;\;\;\;\;\;\;\;\;\;\includegraphics[scale=0.3]{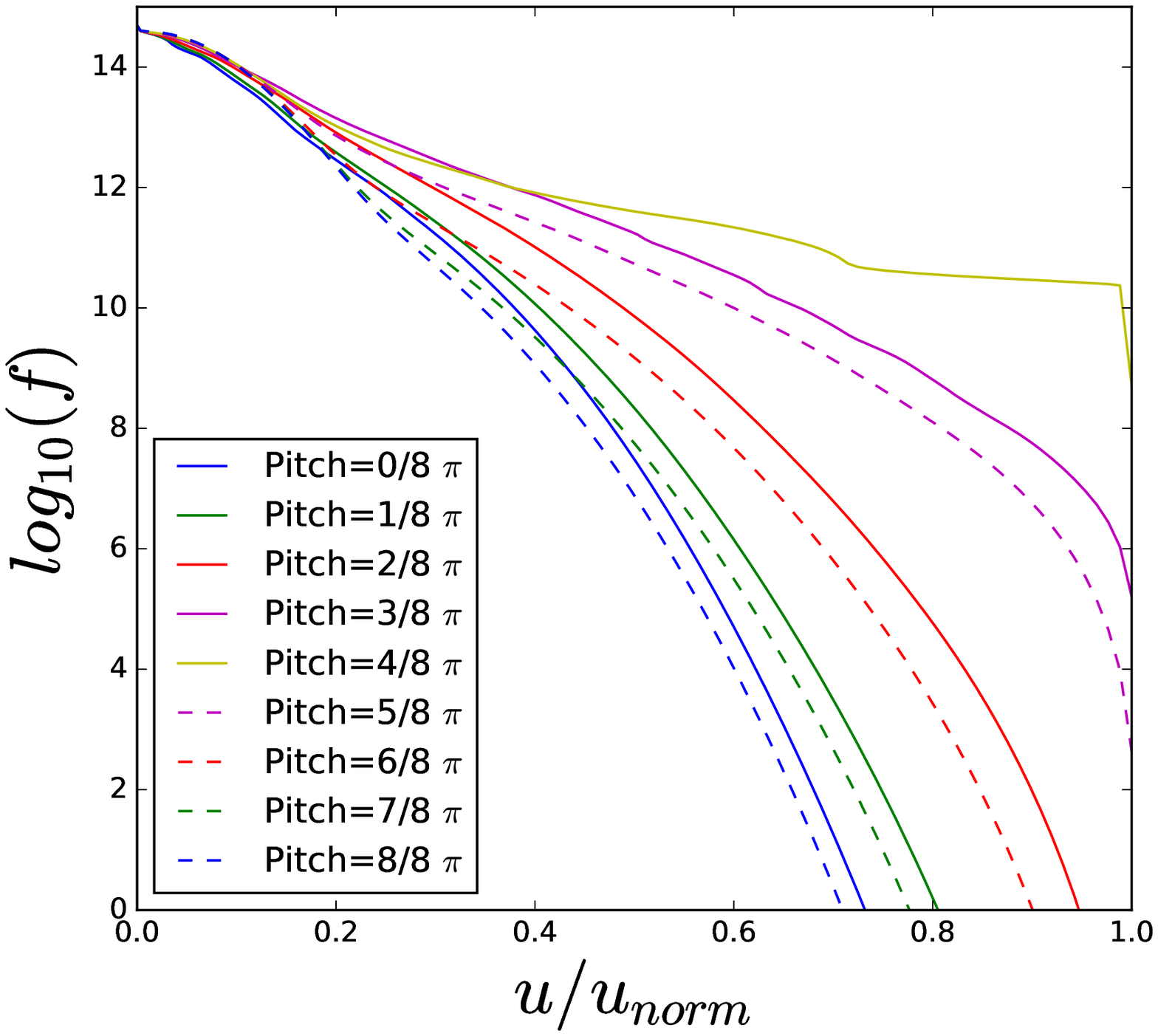}
\caption{Example 2: (a) and (b) are the 2-D contour plots of the distribution function in $(u_{\|0},u_{\perp0})$. (c) and (d) are the 1-D distribution functions in terms of $u$ for several pitch-angles $\vartheta$ at $r/a=0.08$. Figure (a) and (c) are simulated by TORIC-CQL3D and (b) and (d) are simulated by AORSA-CQL3D, where $u_{norm}$ is the momentum corresponding to the energy of He3 4MeV.}
\end{figure*}

\begin{figure*}
(a) \;\;\;\;\;\;\;\;\;\;\;\;\;\;\;\;\;\;\;\;\;\;\;\;\;\;\;\;\;\;\;\;\;\;\;\;\;\;\;\;\;\;\;\;\;\;\;\;\;\;\;\;\;\;\;\;\;\;\;\;\;\;\;\;\;\;\;\;\;\;\;\;(b) \;\;\;\;\;\;\;\;\;\;\;\;\;\;\;\;\;\;\;\;\;\;\;\;\;\;\;\;\;\;\;\;\;\;\;\;\;\;\;\;\;\;\;\;\;\;\;\;\;\;\;\;\;\;\;\;\;\;\;\;\;\;\;\\
\includegraphics[scale=0.20]{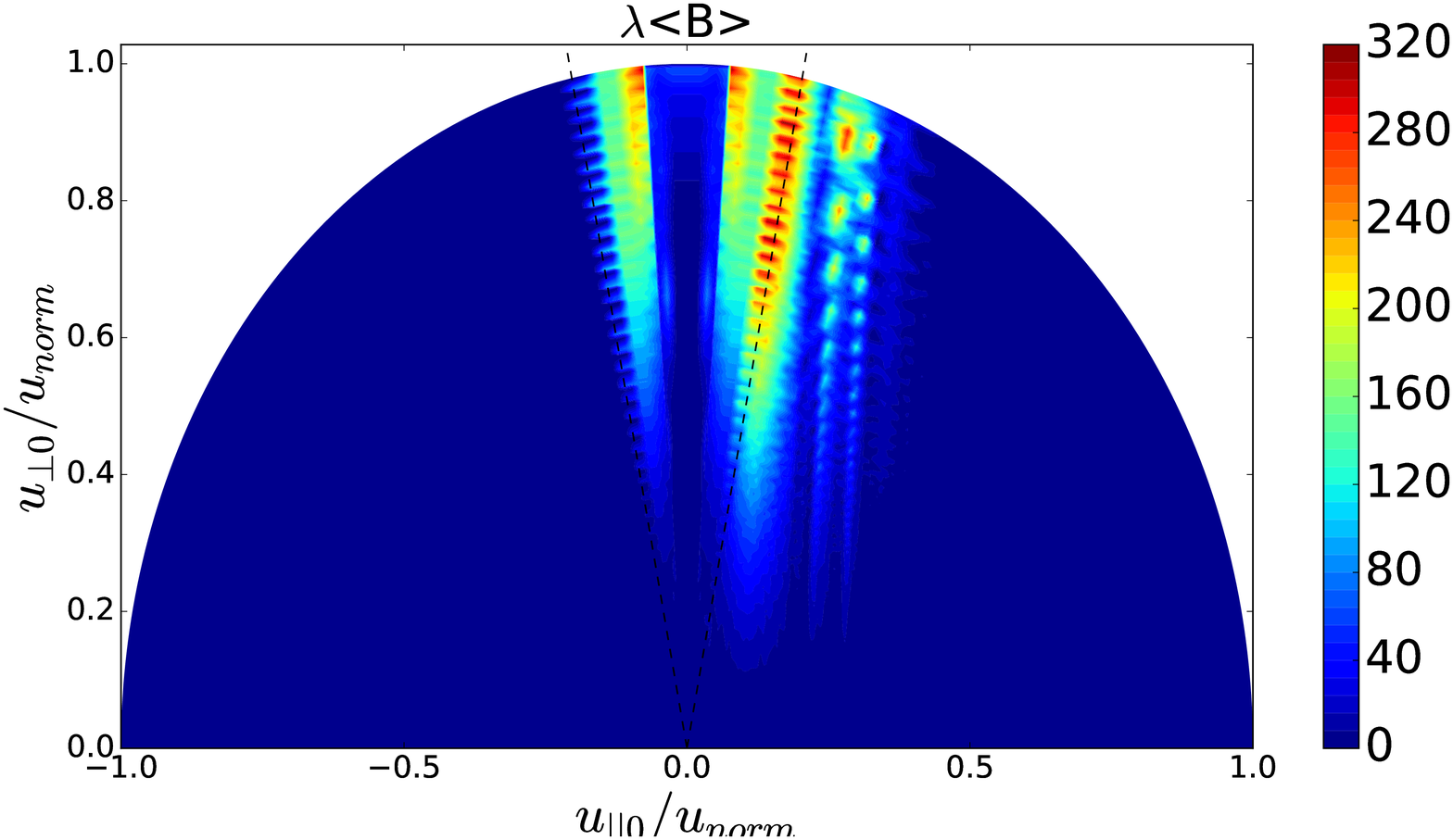}\includegraphics[scale=0.20]{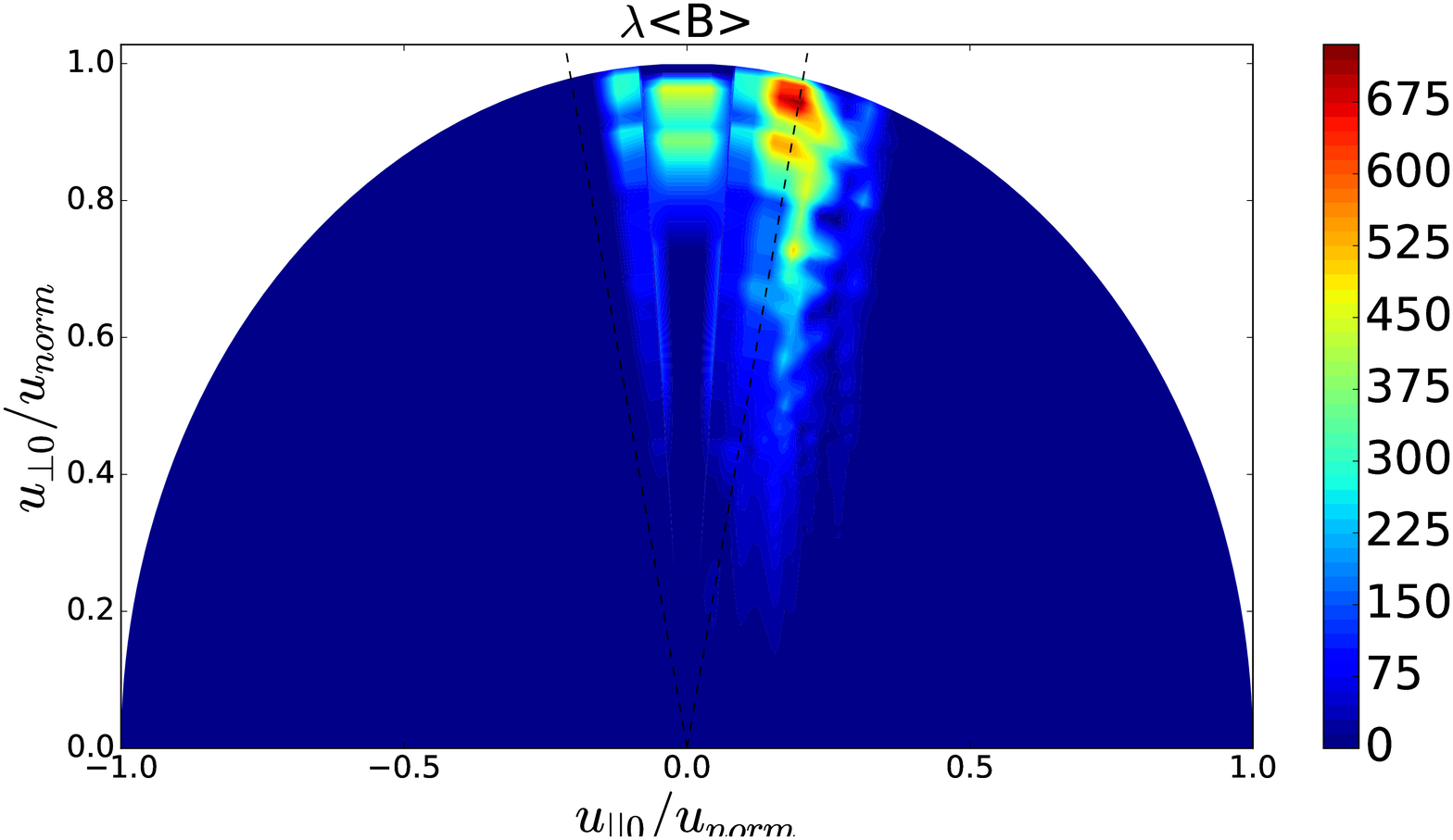}\\
(c) \;\;\;\;\;\;\;\;\;\;\;\;\;\;\;\;\;\;\;\;\;\;\;\;\;\;\;\;\;\;\;\;\;\;\;\;\;\;\;\;\;\;\;\;\;\;\;\;\;\;\;\;\;\;\;\;\;\;\;\;\;\;\;\;\;\;\;\;\;\;\;\;(d) \;\;\;\;\;\;\;\;\;\;\;\;\;\;\;\;\;\;\;\;\;\;\;\;\;\;\;\;\;\;\;\;\;\;\;\;\;\;\;\;\;\;\;\;\;\;\;\;\;\;\;\;\;\;\;\;\;\;\;\;\;\;\;\\
\includegraphics[scale=0.20]{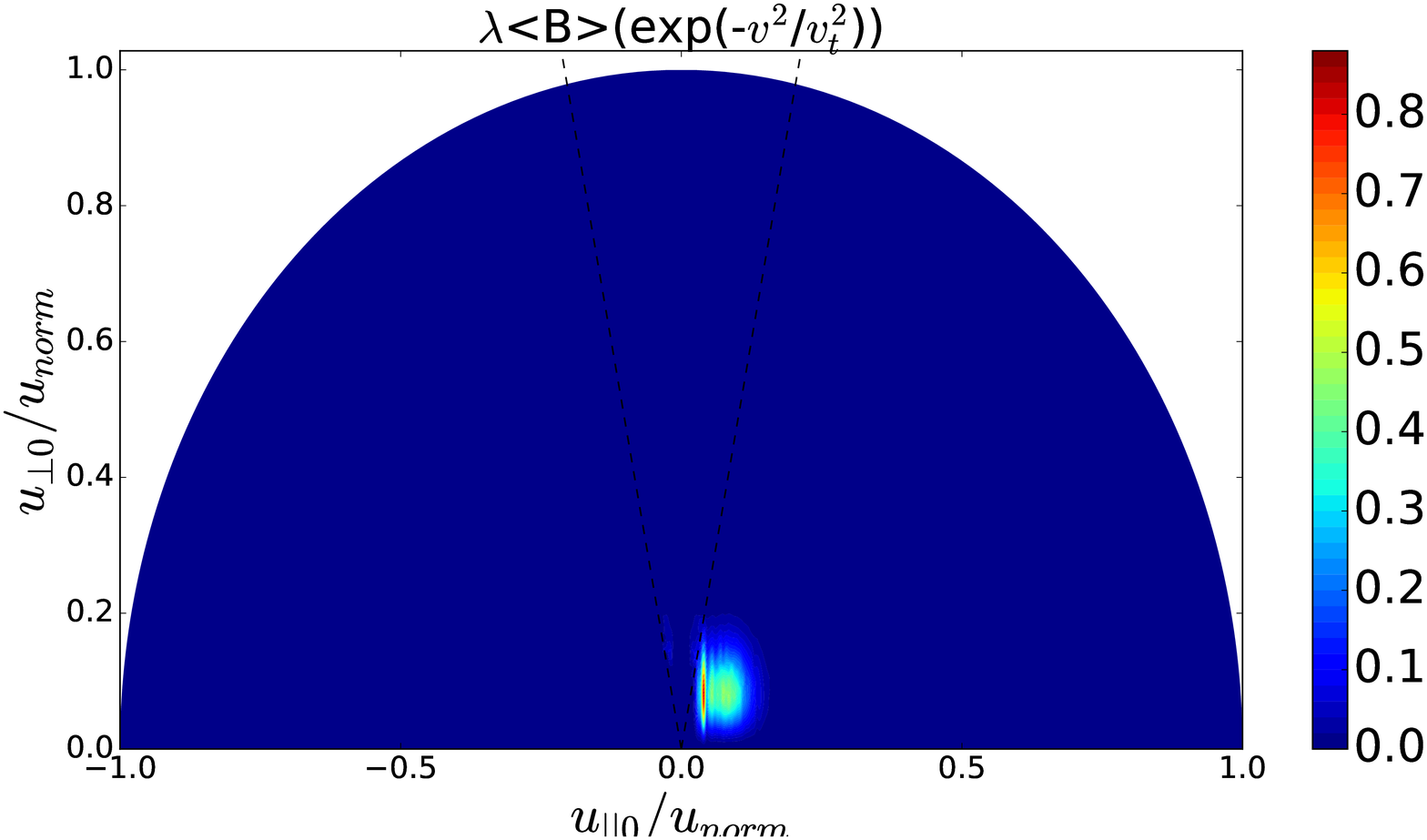}\includegraphics[scale=0.20]{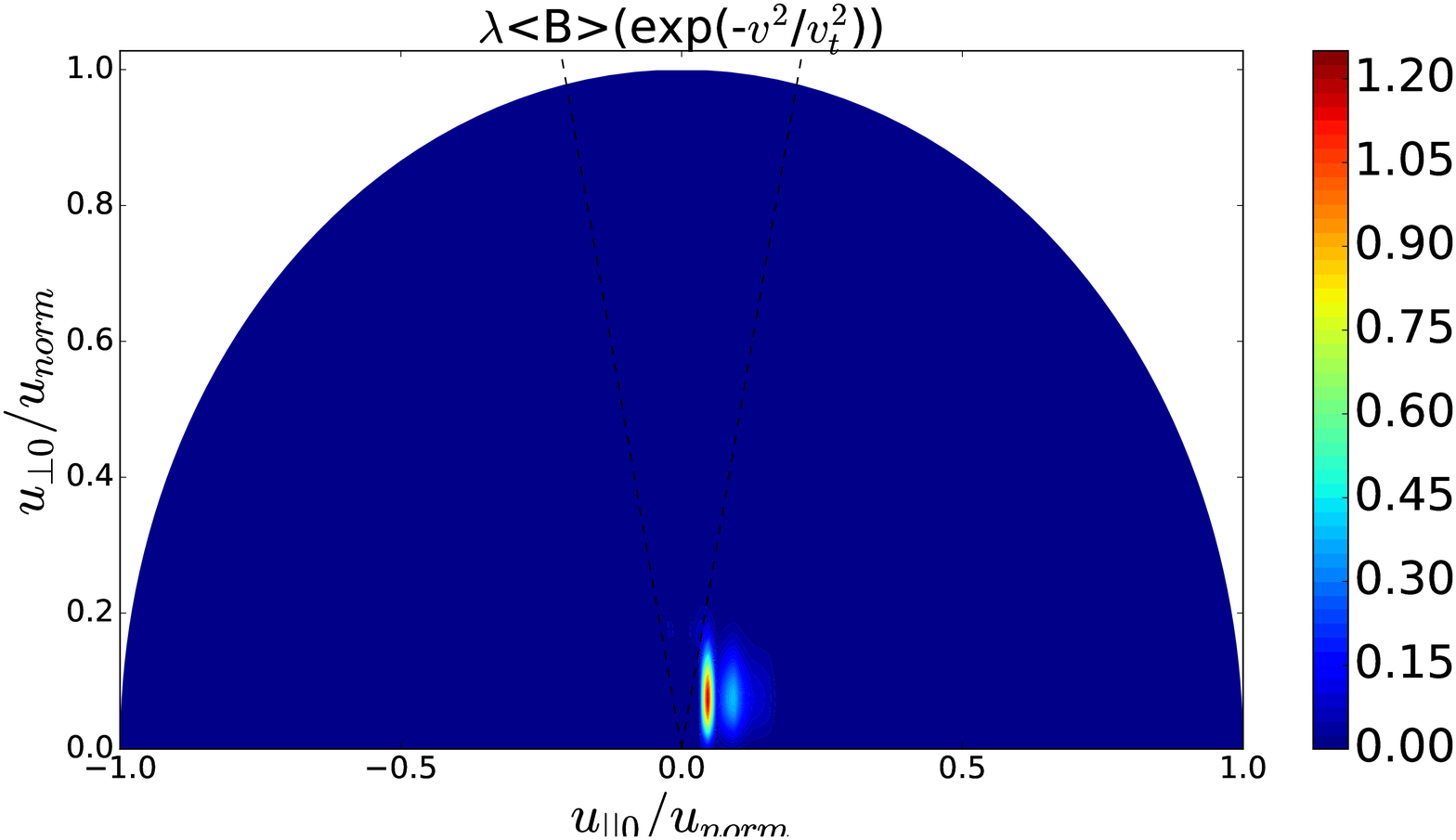}
\caption{Example 2: (a) and (b) are the 2-D contour plots of quasilinear diffusion coefficient $\lambda \langle B\rangle$ in $(u_{\|0},u_{\perp0})$ and (c) and (d) are those weighted by the Maxwellian factor $\lambda \langle B\rangle \exp(-v^2/v_t^2)$ at $r/a=0.08$. Figure (a) and (c) are simulated by TORIC-CQL3D and (b) and (d) are simulated by AORSA-CQL3D, where the contour values are normalized to $u_{norm}$ is the momentum corresponding to the energy of He3 4MeV.} 
\end{figure*}

\section{Discussion}\label{sec:6}
In this paper, we derived and evaluated the quasilinear diffusion coefficients for the reduced model based on the small Larmor radius approximation. Although we present rigorous derivation and proof for the coefficients, the result can be summarized by the two simple statements: (1) use the approximation of Bessel function to the lowest order in $k_\perp\rho_i$ for the coefficient $\textsf{B}$ of Eq. (\ref{Bfull}) in the full model, and (2) use the relations in Eqs. (\ref{BtoC}-\ref{BtoF}) for other coefficients $\textsf{C}$, $\textsf{E}$, and $\textsf{F}$. In other words, it is sufficient for the coefficient $\textsf{B}$ in the reduced model to use $J_0=1$ and $J_1=J_2=0$ for $n=1$ damping and $J_1=k_\perp\rho_i/2$ and $J_2=J_3=0$ for $n=2$ damping. These quasilinear diffusion coefficients guarantee the equivalence with the dielectric tensor in the reduced model of Section \ref{sec:4_2} and the necessary properties of theoretical velocity diffusion such as the wave polarization and the diffusion direction. Because the reduced model is only valid when the Larmor radius is sufficiently small compared to the perpendicular wavelength, it can be inaccurate if there is a significant population of the energetic ions.

In Section \ref{sec:5}, we observe that the diffusion characteristics between the reduced model and the full model are different for the energetic ions with $k_\perp\rho_i \gtrsim 1.0$, although the overall diffusion patterns in the full range of distribution function are similar. The diffusion of the energetic ions by $E_+$ is reduced significantly in the full model because of the decay of Bessel functions in $v_\perp$, while there is no such a decay in the reduced model. Additionally, the dominant polarization changes from $E_+$ to  $E_-$ in the full model, if  $k_\perp\rho_i \gtrsim 1.4$. As shown in the first example in Section \ref{sec:5}, if the wave power density is sufficiently small so that the population of energetic ions with $k_\perp\rho_i \gtrsim 1.0$ is negligible, the self-consistent reduced model is sufficiently accurate compared to the full model. Even for the high power density in the second example, the wave damping profile shows a reasonable agreement with the results of the full model unless there are too many energetic ions with $k_\perp\rho_i \gtrsim 2.0$. 

We also derived the reduced model for the second harmonic $n=2$ damping and implemented it in TORIC-CQL3D. For the high power density of the damping, the error of the reduced model by the FLR approximation is likely larger than the fundamental $n=1$ damping because the approximation of the Bessel function by $J_1\simeq  k_\perp \rho_i/2$ results in the significant difference from the full model for the high $k_\perp\rho_i$.

In high volume tokamaks such as ITER and future reactors, the ICRF wave power density is likely small, and the benefit of the reduced model becomes more important because of the saving of the large computation cost. In this case, the self-consistent quasilinear diffusion coefficients in this paper for the reduced model can be useful, although they are expected to have allowable deviations from the full model. Furthermore, we would like to point out that even for the full model, using the Kennel-Engelmann diffusion coefficients results in the limitations to describe the important physics of the energetic ions in the high power density such as the finite orbit width and the perturbed orbit. The saved computation cost in the reduced model can be used for the extensions of the simulation toward the physically more important directions. For example, it has been noticed the importance of three dimensional simulations that superpose many toroidal modes of two dimensional solutions, time dependent simulations, or the coupling of the core plasma wave with the edge/scrape off layer wave simulations.  

\newpage

\appendix
\section{quasilinear diffusion coefficient in a homogenous magnetic field }\label{sec:A1}
In a uniform magnetic field with spatially uniform plasmas, the Kennel-Engelmann quasilinear diffusion operator  \cite{Kennel:POF1966} can be defined by
\begin{eqnarray}
Q(f)&=&-\frac{q}{m}\left \langle \nabla_v \cdot \left[\left(\mathbf{E} +\frac{\mathbf{v} \times \mathbf{B} }{c}\right) {f} \right] \right \rangle_w \label{Q1}\\
&\simeq&-\frac{q}{m}\nabla_v \cdot \left[ \sum_\mathbf{k} \left \{\stackrel{\leftrightarrow}{I} \left(1-\frac{\mathbf{k} \cdot \mathbf{v}}{\omega} \right) +\frac{\mathbf{k} \mathbf{v}}{\omega} \right\}\cdot\mathbf{E}_\mathbf{-k}   f_{\mathbf{k} } \right], 
 \label{Dql_0}
\end{eqnarray}
where $\stackrel{\leftrightarrow}{I}$ is the unit tensor. We have used the Fourier analyzed fluctuating electric field, $\mathbf{E}=\sum_\mathbf{k}  \mathbf{E}_\mathbf{k}  \exp(i\mathbf{k} \cdot \mathbf{r}-i \omega_\mathbf{k} t)$, the fluctuating magnetic field $\mathbf{B}=\sum_\mathbf{k} \mathbf{B}_\mathbf{k}  \exp(i\mathbf{k} \cdot \mathbf{r}-i \omega_\mathbf{k} t)$, and the fluctuating distribution function, $f=\sum_\mathbf{k} f_\mathbf{k}  \exp(i\mathbf{k} \cdot \mathbf{r}-i \omega_\mathbf{k} t)$. The functions $\mathbf{E}_\mathbf{k} \equiv   \mathbf{E}(\omega_{\mathbf{k} },\mathbf{k})$, $\mathbf{B}_\mathbf{k} \equiv   \mathbf{B}(\omega_{\mathbf{k} },\mathbf{k})$, and $f_\mathbf{k} \equiv  f(\omega_{\mathbf{k} },\mathbf{k})$ satisfy the relation $f_{-\mathbf{k} }\equiv f(\omega_{-\mathbf{k} },-\mathbf{k})=f^*(\omega_{\mathbf{k} },\mathbf{k})$ where $*$ denotes complex conjugate and $\omega_{\mathbf{k} }=-\omega^*_{-\mathbf{k} }$. Faraday's law has been used in going from (\ref{Q1}) to (\ref{Dql_0}) to write $\mathbf{B}_\mathbf{k} =(c/\omega) \mathbf{k} \times \mathbf{E}_\mathbf{k} $. 
The quasilinear operator can be written as 
 \begin{eqnarray}
Q(f)\equiv \frac{q}{m}\left[\frac{1}{v_{\perp} }\frac{\partial}{\partial v_{\perp}} \left(v_{\perp}\Gamma_\perp\right) +\frac{1}{v_{\perp} }\frac{\partial \Gamma_\phi}{\partial \phi} +\frac{\partial \Gamma_\|}{\partial v_{\|}}\right].
 \label{Dql_1}
\end{eqnarray}
The flux in the perpendicular direction is
 \begin{eqnarray}
\Gamma_\perp= -  \sum_\mathbf{k}\bigg \{E^{*}_{\mathbf{k} ,\perp}\bigg(1-\frac{k_{\|}v_{\|}}{\omega} \bigg) +E^{*}_{\mathbf{k} ,\|} \frac{k_{\perp}v_{\|}}{\omega} \cos{(\phi-\beta)} \bigg \}f_{\mathbf{k} }
 \label{Gamma_perp}.
\end{eqnarray}
Here, the velocity is defined as
$\mathbf{v}=v_{\perp} \cos{\phi}\;\mathbf{e_x}+ v_{\perp} \sin{\phi} \;\mathbf{e_y}+v_{\|} \mathbf{e_\|}$, where $\phi$ is the gyro phase angle, and $x$ and $y$ are the orthogonal coordinates in the perpendicular plane to the static magnetic field. The wavenumber vector is defined as $\mathbf{k}=k_{\perp} \cos{\beta}\;\mathbf{e_x}+ k_{\perp} \sin{\beta}\;\mathbf{e_y}+k_{\|}\mathbf{e_\|}$.
The flux in the gyro-phase direction is
 \begin{eqnarray}
\Gamma_\phi&=& -  \sum_\mathbf{k}\bigg\{E^{*}_{\mathbf{k} ,\phi} \bigg(1-\frac{k_{\perp}v_{\perp}}{\omega} \cos{(\phi-\beta)}-\frac{k_{\|}v_{\|}}{\omega} \bigg)\nonumber \\
&-& E^{*}_{\mathbf{k} ,\perp}\frac{k_{\perp}v_{\perp}}{\omega}  \sin{(\phi-\beta)} -E^{*}_{\mathbf{k} ,\|} \frac{k_{\perp}v_{\|}}{\omega} \sin{(\phi-\beta)} \bigg \}{f}_\mathbf{k} ,
 \label{Gamma_alpha}
\end{eqnarray}
and the flux in the parallel direction is
 \begin{eqnarray}
\Gamma_\|=  - \sum_\mathbf{k}\bigg\{E^{*}_{\mathbf{k} ,\|} \bigg(1-\frac{k_{\perp}v_{\perp}}{\omega} \cos{(\phi-\beta)} \bigg) + E^{*}_{\mathbf{k} ,\perp}\frac{k_{\|}v_{\perp}}{\omega}\bigg \}{f}_\mathbf{k} .
 \label{Gamma_par}
\end{eqnarray}
Here, the perturbed fluctuating distribution function consistent with a single mode wave is
 \begin{eqnarray}
f_\mathbf{k}  &=& - \frac{q}{m}e^{-i \mathbf{k} \cdot \mathbf{r} + i \omega t} \int^t_{-\infty} d t^\prime e^{i \mathbf{k} \cdot \mathbf{r}^{\prime} -i \omega t^{\prime}} \mathbf{E_\mathbf{k} } \cdot \left[\stackrel{\leftrightarrow}{I} \left(1-\frac{\mathbf{v}^\prime \cdot \mathbf{k}}{\omega} \right) +\frac{\mathbf{v}^\prime \mathbf{k}}{\omega} \right] \cdot  \nabla_{v^{\prime}} f ,  \label{f_k1}
\end{eqnarray}
where $(t^{\prime},\mathbf{r}^{\prime},\mathbf{v}^\prime)$ is a point of phase space along the zero-order particle trajectory. The trajectory end point corresponds to $(t,\mathbf{r},\mathbf{v})$. The background distribution, $f=f(t,\mathbf{r},v_{\perp}, v_{\|})$, is gyro-phase independent because of the fast gyro-motion. As a result,
 \begin{eqnarray}	
f_\mathbf{k}  &=&-\frac{q}{m} \int_0^{\infty} d\tau \exp({i\alpha})  \bigg\{ \cos{(\eta+\Omega \tau)}(( E_{\mathbf{k} ,+}+E_{\mathbf{k} ,-})U-E_{\mathbf{k},\|}V)\nonumber\\
&&-i\sin{(\eta+\Omega \tau)}( E_{\mathbf{k} ,+}-E_{\mathbf{k} ,-})U +E_{\mathbf{k} ,\|} \frac{\partial f}{\partial v_{\|}}\bigg \}. \label{f_k2}
\end{eqnarray}
Here, $\tau=t-t^{\prime}$, and $\alpha=(\omega-k_{\|}v_{\|})\tau -\lambda(  \sin{(\eta+\Omega \tau)}- \sin{(\eta)}) $, where $\lambda= {k_{\perp}v_{\perp}}/{\Omega}$ and $\eta=\phi-\beta$. Also, $U={\partial f}/{\partial v_{\perp}}+ ({k_{\|}}/{\omega})\left( v_{\perp}{\partial f}/{\partial v_{\|}}-v_{\|}{\partial f}/{\partial v_{\perp}}\right)$, and $V= ({k_{\perp}}/{\omega})\left( v_{\perp}{\partial f}/{\partial v_{\|}}-v_{\|}{\partial f}/{\partial v_{\perp}}\right)$. We follow Stix' notation \cite{Stix:AIP1992}.

For the energy transfer, the contribution of the flux in the gyro-phase direction vanishes due to the integral over $\phi$. Using the Bessel function expansion for the sinusoid phase, 
 \begin{eqnarray}
e^{i\lambda \sin {\eta}}&=&\sum_n e^{in\eta} J_n(\lambda),\nonumber\\
\sin{\eta} e^{i\lambda \sin {\eta}}&=&-\sum_n i e^{in\eta} J_n^{\prime}(\lambda),\nonumber\\
\cos{\eta} e^{i\lambda \sin {\eta}}&=&\sum_n \frac{n}{\lambda} e^{in\eta} J_n(\lambda),\label{Bess00}
\end{eqnarray}
the gyro-averaged quasilinear diffusion  \cite{Kennel:POF1966,Stix:AIP1992} is
\begin{eqnarray}
Q(f)&=&\frac{{\pi} q^2 }{m^2}  \int_{-\infty} ^{\infty}   \sum_n G \bigg (v^2_{\perp} \delta (\omega-k_{\|}v_{\|}-n\Omega) |\chi_{\mathbf{k},n}|^2 G(f) \bigg ) \label{P_abs1}\label{P_abs2}
\end{eqnarray} 
where $\chi_{\mathbf{k} ,n}= E_{\mathbf{k} ,+} J_{n-1}/\sqrt{2} +E_{\mathbf{k} ,-} J_{n+1}/\sqrt{2}+({v_{\|}}/v_{\perp})E_{\mathbf{k} ,\|} J_{n}$ is the effective electric field, and the operator $G$ is 
\begin{eqnarray}
G(f)=\left(1-\frac{k_{\|}v_{\|}}{\omega}\right)\frac{1}{v_{\perp}}\frac{\partial f}{\partial v_{\perp}}+\frac{k_{\|}v_{\perp}}{\omega}\frac{1}{v_{\perp}}\frac{\partial f}{\partial v_{\|}} \label{KE_diff}
\end{eqnarray} 

\clearpage
\newpage

\section*{Acknowledgments}
 This work was supported by US DoE Contract No. DE-FC02-01ER54648 under a Scientific Discovery through
Advanced Computing Initiative. NB and EV were supported by the US DOE under DE-AC02-CH0911466. This research used resources of MIT, PPPL, and NERSC by the US DoE Contract No.DE-AC02-05CH11231

\newpage


\end{document}